%% file: main.tex
\documentclass[acmtog,nonacm]{acmart}
\usepackage{booktabs} %
\usepackage{enumitem}
\usepackage{pifont} %
\usepackage{threeparttable} %
\usepackage{siunitx} %
\usepackage{multicol}

\newcommand{\uname}{HNLPU}

\usepackage{multirow}

\usepackage{soul}
\soulregister\cite7
\soulregister\ref7

\usepackage{caption}%

  \newcommand*{\revadd}[1]{{#1}}  
  \newcommand*{\revdel}[1]{}  
  \newcommand*{\revlink}[1]{}
  \newcommand*{\revtarget}[1]{}

\usepackage{etoolbox}
\begin{document}

\title{Hardwired-Neurons Language Processing Units as General-Purpose Cognitive Substrates}

\author{Yang Liu\textsuperscript{1}\textsuperscript{2}\textsuperscript{*}} 
\author{Yi Chen\textsuperscript{1}\textsuperscript{3}\textsuperscript{*}} 
\author{Yongwei Zhao\textsuperscript{1}} 
\author{Yifan Hao\textsuperscript{1}} 
\author{Zifu Zheng\textsuperscript{1}\textsuperscript{2}} 
\author{Weihao Kong\textsuperscript{1}\textsuperscript{2}} 
\author{Zhangmai Li\textsuperscript{1}\textsuperscript{3}} 
\author{Dongchen Jiang\textsuperscript{1}\textsuperscript{2}} 
\author{Ruiyang Xia\textsuperscript{1}\textsuperscript{2}} 
\author{Zhihong Ma\textsuperscript{1}\textsuperscript{2}} 
\author{Zisheng Liu\textsuperscript{1}\textsuperscript{2}} 
\author{Zhaoyong Wan\textsuperscript{1}\textsuperscript{3}} 
\author{Yunqi Lu\textsuperscript{1}\textsuperscript{2}} 
\author{Ximing Liu\textsuperscript{1}\textsuperscript{2}} 
\author{Hongrui Guo\textsuperscript{1}\textsuperscript{2}} 
\author{Zhihao Yang\textsuperscript{4}\textsuperscript{2}} 
\author{Zhe Wang\textsuperscript{1}\textsuperscript{2}} 
\author{Tianrui Ma\textsuperscript{1}} 
\author{Mo Zou\textsuperscript{1}} 
\author{Rui Zhang\textsuperscript{1}} 
\author{Ling Li\textsuperscript{4}} 
\author{Xing Hu\textsuperscript{1}} 
\author{Zidong Du\textsuperscript{1}} 
\author{Zhiwei Xu\textsuperscript{1}} 
\author{Qi Guo\textsuperscript{1}\textsuperscript{\dag}}
\author{Tianshi Chen\textsuperscript{5}} 
\author{Yunji Chen\textsuperscript{1}\textsuperscript{2}}

\affiliation{
  \institution{\textsuperscript{1}State Key Lab of Processors, Institute of Computing Technology, CAS}
  \country{China}
}

\affiliation{
  \institution{\textsuperscript{2}University of Chinese Academy of Sciences}
  \country{China}
}

\affiliation{
  \institution{\textsuperscript{3}University of Science and Technology of China}
  \country{China}
}

\affiliation{
  \institution{\textsuperscript{4}Institute of Software, CAS}
  \country{China}
}

\affiliation{
  \institution{\textsuperscript{5}Cambricon Technologies}
  \country{China}
}

\email{{liuyang22z1, zhaoyongwei, haoyifan, zhengzifu24s, kongweihao21b,
jiangdongchen23s, xiaruiyang, mazhihong24z, liuzisheng24s, luyunqi24s,
liuximing22z, guohongrui21b, wangzhe24s, matianrui, zoumo, zhangrui, huxing,
duzidong, zxu, guoqi, cyj}@ict.ac.cn, {chen\_yi, lizhangmai, wanzhy}@mail.ustc.edu.cn, {yangzhihao2021, liling}@iscas.ac.cn, tchen@cambricon.com}

\thanks{\dag Corresponding Author: Qi Guo \\ *Contribute Equally}

\renewcommand{\shortauthors}{Liu et al.}

\input{tex/abstract}

\maketitle

\input{tex/introduction}

\input{tex/background}

\input{tex/design}

\input{tex/architecture}

\input{tex/dataflow}

\input{tex/methodology}
\input{tex/experiments}

\input{tex/related}
\input{tex/conclusion}

\bibliographystyle{ACM-Reference-Format}

\bibliography{hn}

\appendix

\input{tex/dataflow_appendix}

\input{tex/cost_analysis}

\end{document}

%% file: tex/abstract.tex
\begin{abstract}

The rapid advancement of Large Language Models (LLMs) has established language as a core general-purpose cognitive substrate, driving the demand for specialized Language Processing Units (LPUs) tailored for LLM inference. 
To overcome the growing energy consumption of LLM inference systems,
this paper proposes a Hardwired-Neurons Language Processing Unit (\uname{}), which physically hardwires LLM weight parameters into the computational fabric,
achieving several orders of magnitude computational efficiency improvement by extreme specialization.
However, a significant challenge still lies in the scale of modern LLMs. 
\revadd{A straightforward} hardwiring \revadd{of} \textsc{gpt-oss} 120\,B \revadd{would require} fabricating \revadd{photomask sets valued at over} 6 billion dollars,
rendering \revadd{this} straightforward solution economically impractical.

Addressing this challenge, we propose the novel \emph{Metal-Embedding} methodology. 
Instead of embedding weights in a 2D grid of silicon device cells, Metal-Embedding embeds weight parameters into the 3D topology of metal wires.
This brings two benefits: (1) a 15× increase in density, and (2) 60 out of 70 \revadd{photomask layers are} homogeneous across chips, including all EUV photomasks.
In total, Metal-Embedding reduced the photomask cost by 112×, bringing the Non-Recurring Engineering (NRE) cost of \uname{} into an economically viable range.
Experimental results show that \uname{} achieved 249,960 tokens/s (5,555×/85× \revadd{that} of GPU/WSE), 36 tokens/J
(1,047×/283× \revadd{that} of GPU/WSE),
13,232\,mm\textsuperscript{2} total die area, \revadd{\$\,59.46\,M--123.5\,M} estimated NRE at 5\,nm technology.
Analysis shows that \revadd{\uname{} achieved 41.7--80.4× improvement in cost-effectiveness and 357× reduction in carbon footprint compared to OpenAI-scale H100 clusters, under an annual weight updating assumption.}

\end{abstract}

%% file: tex/introduction.tex
\section{Introduction}
\label{sec::introduction}

\revadd{Current} trends in artificial intelligence development indicate that language capabilities serve as the core foundation for constructing advanced cognitive and learning systems. 
This is exemplified by the rapid advancement of Large Language Models (LLMs), which have achieved technological unification in Natural Language Processing (NLP)~\cite{attention, gpt3, emergent}.
By demonstrating human-level performance across diverse downstream tasks, LLMs are expanding their application scope while exhibiting a growing trend toward enhanced generality---enabling a single model to handle multiple tasks that previously required distinct systems.

Driven by widespread adoption \revadd{despite the} heavy hardware overhead of LLMs, there has been a surge in demand for dedicated processors specifically designed for LLM inference, collectively known as Language Processing Units (LPUs).
For instance, Groq LPU~\cite{groq-lpu1, groq-lpu2} and Cerebras WSE~\cite{cerebras2021, cerebras2023} pre-load model weights into on-chip SRAM, while Etched Sohu~\cite{etched_sohu} hardens the transformer dataflow into its compute fabric.
These LPUs leverage the advanced specialization targeting LLMs, achieving $4\sim20\times$ energy efficiency compared to traditional CPUs, GPUs, and NPUs.

\begin{figure}[t]
  \centering
  \includegraphics[width=\columnwidth]{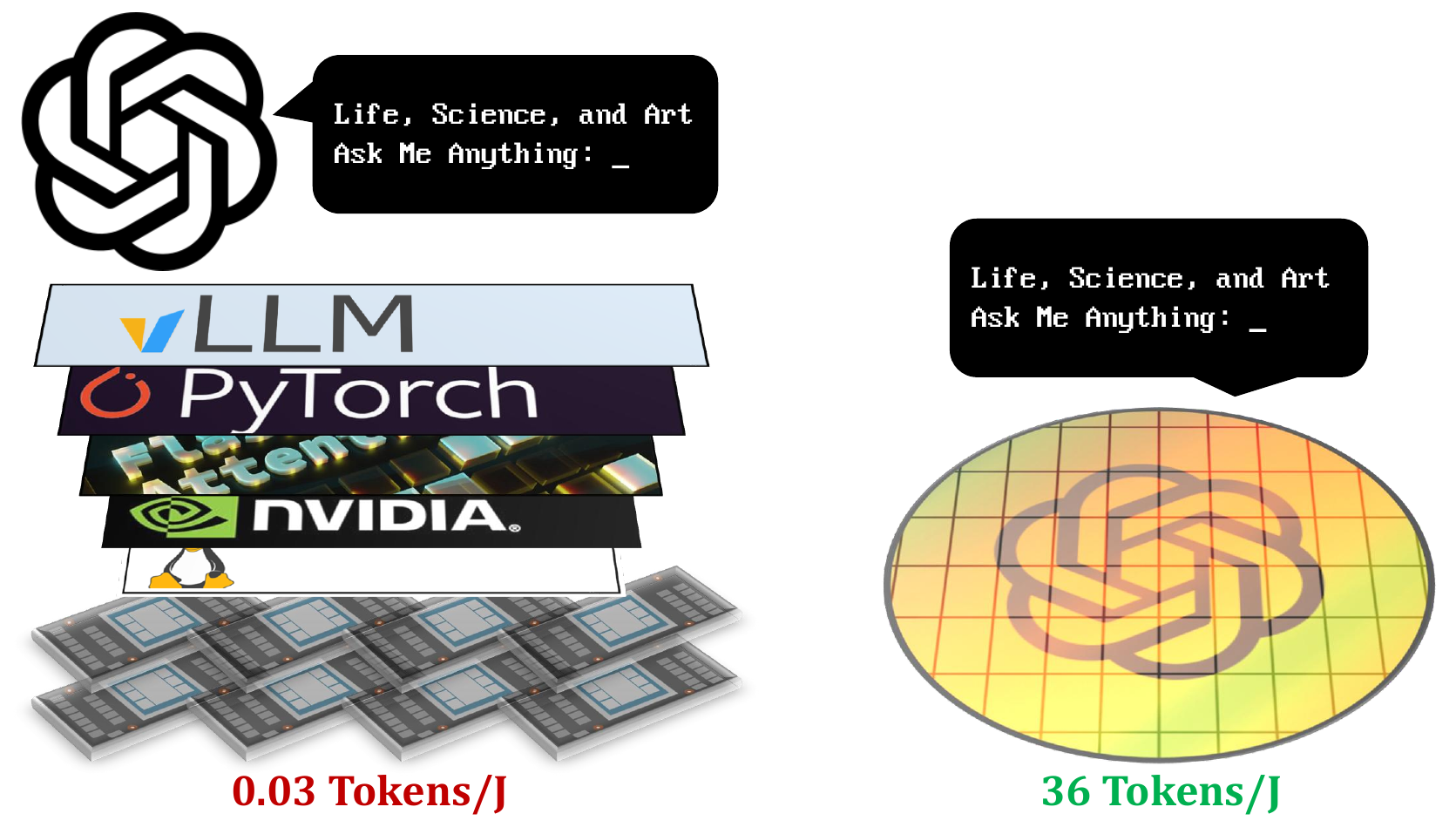}
  \caption{\textbf{\revadd{Hardwired LPU} as a general-purpose processor.} To the left: AI Infrastructures are originated in the rapidly evolving deep learning which appreciates universality over extreme efficiency. To the right: As LLM develops, the responsibilities of universality are shifting from HW/SW to LLMs. An extremely specialized Hardwired LPU can also be helpful in general tasks.}
  \label{fig:intro}
  \vspace{-1em}
\end{figure}

Although LPUs have brought a certain improvement in hardware efficiency, AI infrastructure still suffers from tremendous energy \revadd{costs}. 
For example, AI datacenters are estimated to occupy 12\% of the total electricity capacity in the US \revadd{by} 2028, which is clearly unsustainable~\cite{energy_cost}.
This is due to the fact that current LPUs and GPUs still see models as dynamically changing data, despite the fact that leading enterprises typically deploy only a single or a few proprietary \revadd{LLMs}.
The hundreds of billions of weight parameters are repeatedly fetched during each autoregressive decoding step, consuming the majority of system power.

To fundamentally solve the limitation, we argue \revadd{for pushing} the specialization of LPU to the extreme. By physically hardwiring weight parameters into the LPU fabric, \revadd{Hardwired LPU} can achieve perfect architecture-model matching, zero parameter fetching overhead, and extreme computational efficiency from constants-arithmetic circuits.
As one dominant pre-trained LLM can serve as a general-purpose cognitive substrate for a wide variety of tasks, the extreme specialization once considered too inflexible is transforming into a reasonable choice with huge potential benefits (Figure~\ref{fig:intro}).

However, previous attempts on \revadd{hardwiring} would be stopped by the extraordinarily large scale of modern LLM.
The most optimistic estimation on hardwiring \textsc{gpt-oss} 120\,B~\cite{openai_gpt_oss_2025} would require \revadd{a} 176,000\,mm\textsuperscript{2} \revadd{constant-multiply-and-accumulate} units (CMAC) array at 5\,nm technology.
Unlike previous wafer-scale practices \revadd{that step} a repeated lithographic pattern on the wafer, here the patterns are heterogeneous everywhere, because embedded weight parameters are different everywhere.
The photomask sets \revadd{are} valued \revadd{at} over \emph{6 billion dollars}, rendering the straightforward implementation of \revadd{Hardwired LPU} economically prohibitive.

\revadd{An} economically viable \revadd{Hardwired LPU requires a breakthrough in} weight-embedding methodology.
In this paper, we propose the Metal-Embedding (ME) methodology, achieving multiple orders-of-magnitude reduction on photomask counts required to implement \revadd{Hardwired LPU}.
Roughly speaking, instead of embedding weights in the 2D grid of cells (either MAC/CMAC/SRAM/ReRAM cells), ME embeds weights in the 3D topology of metal wires (Figure~\ref{fig:metal-embedding}).
The benefits of ME are two-fold.
First, due to the inherently richer expression of 3D structure, ME reduced the area by an order-of-magnitude (-93.4\%) compared with the CMAC grid.
Second, these metal wires can be placed at higher metal layers (M8+), to be cost-efficiently fabricated with trailing-edge optical lithography (193i).
As a result, ME keeps photomask homogeneous (shared) across all chips for all layers in Front-End-of-Line (FEOL) and critical layers in Back-End-of-Line (BEOL), including all Extreme Ultraviolet (EUV) photomask in the process.
This reduced the total cost on photomask by another order of magnitude: -86.5\% for initial tapeout, -92.3\% for parameter-only update re-spin.

We evaluate \revadd{the first Hardwired LPU design} \uname{} at 5\,nm technology based on post-layout characteristics, and compare with baselines (NVIDIA H100 GPU, Cerebras WSE-3) at the same technology node.
Experimental results demonstrate that \uname{} achieved 249,960 tokens/s throughput (5,555× of H100, 85× of WSE-3),
36 tokens/J energy efficiency (1,047× of H100, 283× of WSE-3),
at 13,232\,mm\textsuperscript{2} total die area divided into 16 chips.
Estimated Non-Recurring Engineering cost (NRE) of \uname{} is \revadd{\$\,59.46\,M--123.5\,M} for the initial tapeout, \revadd{\$\,18.53\,M--37.06\,M} for a parameter-only update re-spin. 
\revadd{Under the OpenAI-scale deployment, the 3-Year Total Cost of Ownership (TCO) is lowered by 41.7--80.4× compared to H100 clusters with annual parameter updates under the OpenAI-scale volume,
and the total carbon footprint (tCO\textsubscript{2}e) is lowered by 357×.}

This paper makes the following contributions:

\begin{itemize}
    \item %
    We explored the concept and feasibility of \revadd{Hardwired LPU} for the first time.
    \item %
    We propose the Metal-Embedding methodology, reduced the photomask cost of \revadd{Hardwired LPU} by two orders of magnitude.
    \item %
    We detailed the architecture and dataflow of the first \revadd{Hardwired LPU} design, \uname{}. It implemented a \textsc{gpt-oss} 120\,B (FP4) in 13,232\,mm\textsuperscript{2} total die area, achieving unprecedented throughput and energy efficiency.
    \item %
    We estimated the NRE, TCO, and tCO\textsubscript{2}e of \uname{} to show its strong economical and environmental advantage for typical cloud serving scenarios.
\end{itemize}
\vspace{-1em}

%% file: tex/background.tex
\section{Background and Motivation}
\label{sec::background}

\revdel{Current AI development trends underscore language as the foundational substrate for constructing advanced cognitive capabilities. This centrality is supported by three key arguments. First, language is demonstrably critical to human intelligence. Neurocognitive research reveals strong correlations between linguistic competence and higher-order cognitive abilities, such as abstract reasoning~\cite{lupyan2015words}. These findings strongly suggest that language should similarly form the core of advanced artificial intelligence. Second, the sheer abundance of textual data vastly exceeds that available in other modalities. For instance, Common Crawl~\cite{commoncrawl_overview} has amassed an 8\,PB text corpus from web pages, growing consistently by approximately 250\,TB per month. This unparalleled scale of pre-training data is instrumental in enabling the zero-shot generalization abilities observed in Large Language Models (LLMs). Third, language constitutes the most natural and universal interface for human-AI collaboration. Robust language understanding and generation capabilities are indispensable for AI to integrate into human workflows and societal structures.}

\revdel{The rise of LLMs has further cemented language's foundational status through technical unification. Historically, NLP was fragmented into specialized subfields (e.g., sentiment analysis, machine translation, named entity recognition), each requiring task-specific systems. Today, a single pre-trained LLM achieves state-of-the-art performance across common NLP tasks. This convergence stems from scaling model parameters and training data, enabling one general-purpose model to outperform prior specialized approaches. Consequently, LLMs now serve as universal engines for language-powered cognition, spanning reasoning, code generation, and multimodal interaction through linguistic interfaces.}

\revdel{This unification demonstrates that a single, highly capable LLM can function as a general-purpose cognitive substrate. While multiple implementations exist (e.g., GPT, Claude, Gemini, Qwen and DeepSeek), the focus of competition is shifting from capabilities to computational efficiency, a domain where model-architecture co-design becomes decisive.}
\revdel{The demonstrated generality of modern LLMs makes architectural specialization strategically viable. For instance, models like \textsc{gpt-oss} 120\,B which achieves state-of-the-art performance, exemplify how a singular, optimally scaled Transformer architecture can serve as a universal language engine. By fixing such a model as a standardized substrate, unprecedented hardware optimizations become feasible, while maintaining full generality through the model's inherent language-based cognitive flexibility.}

\revdel{The emergence of LLM as universal cognitive engines has intensified demand for dedicated inference hardware.
We define processors explicitly optimized for language models as Language Processing Units (LPUs), a complementary paradigm following CPUs/GPUs/NPUs.
However, currently LPUs are predominantly repurposed from general-purpose processors.
For example, the most widely-deployed NVIDIA Hopper architecture incorporates dedicated Transformer Engines, but still retains the architectural lineage of its Tensor Core GPUs originated before LLMs.
These repurposed processors suffer from efficiency limitations intrinsic to their origin in more general applications, known as "Turing Tariffs"~\cite{kelly2020turing}.}

\revdel{The major efficiency limitation in current architectures is two-fold.
First, the architecture-model mismatch fails to accommodate divergent computational demands across inference phases (e.g., compute-bound prefill vs. memory-bound decoding), forcing industrial compromises like prefill/decode and attention/feed-forward decoupling that deploy heterogeneous hardware subsystems specifically~\cite{deepspeed, step3}.
Second, repurposed processors dynamically load models as ever-changing data, despite the fact that enterprises typically deploy only a single or a few proprietary LLM. The $>$100\,B model parameters are repeatedly fetched during each autoregressive decoding step, dominating the whole system power.}

\revdel{Emerging architectures like Cerebras Wafer-Scale Engine (WSE) and Groq LPU attempt to circumvent these bottlenecks through deterministic scheduling and massive on-chip SRAM integration, reportedly achieved up to 20× energy efficiency gain over GPUs~\cite{groq-lpu1, cerebras2023}.
These practices demonstrated the great benefits and viability of pushing the specialization of LPU hardware.
However, Turing Tariffs still exist in these architectures. Only on one hand, parameter fetching still dominates the power consumption as the SRAM access cost is typically several times higher than computation~\cite{sze2017efficient}.}

\revdel{To fundamentally address these limitations, we propose to push the specialization of LPU to an architectural extreme: permanent weight hardwiring within LPU chips.
By physically embedding model parameters into the fabric of computing units, $\sim$1000× efficiency gain could be obtained from perfect architectural matching, zero parameter fetching overhead, and constant arithmetic.
Such model-specialized architecture aligns with the industrial consolidation trend, where enterprises increasingly standardize on one dominant LLM architecture, transforming what was once considered an inflexibility into a fairly reasonable choice with huge potential benefits.}

{
\subsection{Hardwired Language Processing Units}\revtarget{sec21}
}

\revadd{In principle, orders-of-magnitude gains in computational efficiency can be achieved by directly fabricating neural network models into hardware. Since the 1980s, hardwired neural networks have been implemented in VLSI~\cite{vlsinn}, optical~\cite{optic1,optic2}, and \revtarget{cite1}printed flexible~\cite{mubarik2020printed,weller2021printed,ozer2023malodour,ozer2020hardwired,chakraborty2024hasics} circuits. However, the diversity and rapid evolution of neural networks have rendered such hardwired implementations of limited practical value and prone to rapid obsolescence. As a result, instead of hardwired neural networks, general-purpose processors such as GPUs and NPUs have prevailed due to their programmability. Contemporary AI infrastructure and software stacks including programming languages (e.g., CUDA, Triton), libraries (e.g., cuDNN, flash-attn), frameworks (e.g., PyTorch, vLLM) are built on a sweet spot between computational efficiency and architectural generality.}

\revadd{The rise of large language models (LLMs) has fundamentally altered this landscape. Unlike earlier task-specific neural networks designed for isolated tasks, LLMs function as general-purpose cognitive substrates capable of reasoning, planning, and interacting across a wide range of domains.
\textbf{For the first time in history, we possess a single neural network model that is valuable enough to justify hardwired implementation.}}

\revadd{Although technically \emph{hardwired} (in the sense that its parameter weights are physically immutable), the circuit embodies a new paradigm of the \emph{general-purpose} processor.
We refer to this novel processor paradigm as the Hardwired Language Processing Unit (Hardwired LPU).
By leveraging the emergent capabilities of in-context learning and zero-shot reasoning, a hardwired LPU treats natural language prompts as a high-level instruction stream, effectively replacing the traditional binary Instruction Set Architecture (ISA) with a semantic interface.
Consequently, the user programs the Hardwired LPU not by altering its models and weights, but by prompting with tokens to perform arbitrary tasks.}

\revadd{%
In this paper, we demonstrate that a single-node Hardwired LPU can outperform a middle-sized GPU cluster without requiring expensive software support, thus establishing a decisive operational-expenditure (OpEx) advantage over challengers. Therefore, we envision Hardwired LPUs as game-changing platforms for long-term, high-volume deployment, while GPU clusters assume the role of short-cycle model development and evaluation testbeds.}

{
\subsection{Economic Challenges}}\revtarget{sec22}

\begin{figure}\revtarget{fig2}
    \centering
    \includegraphics[width=\columnwidth]{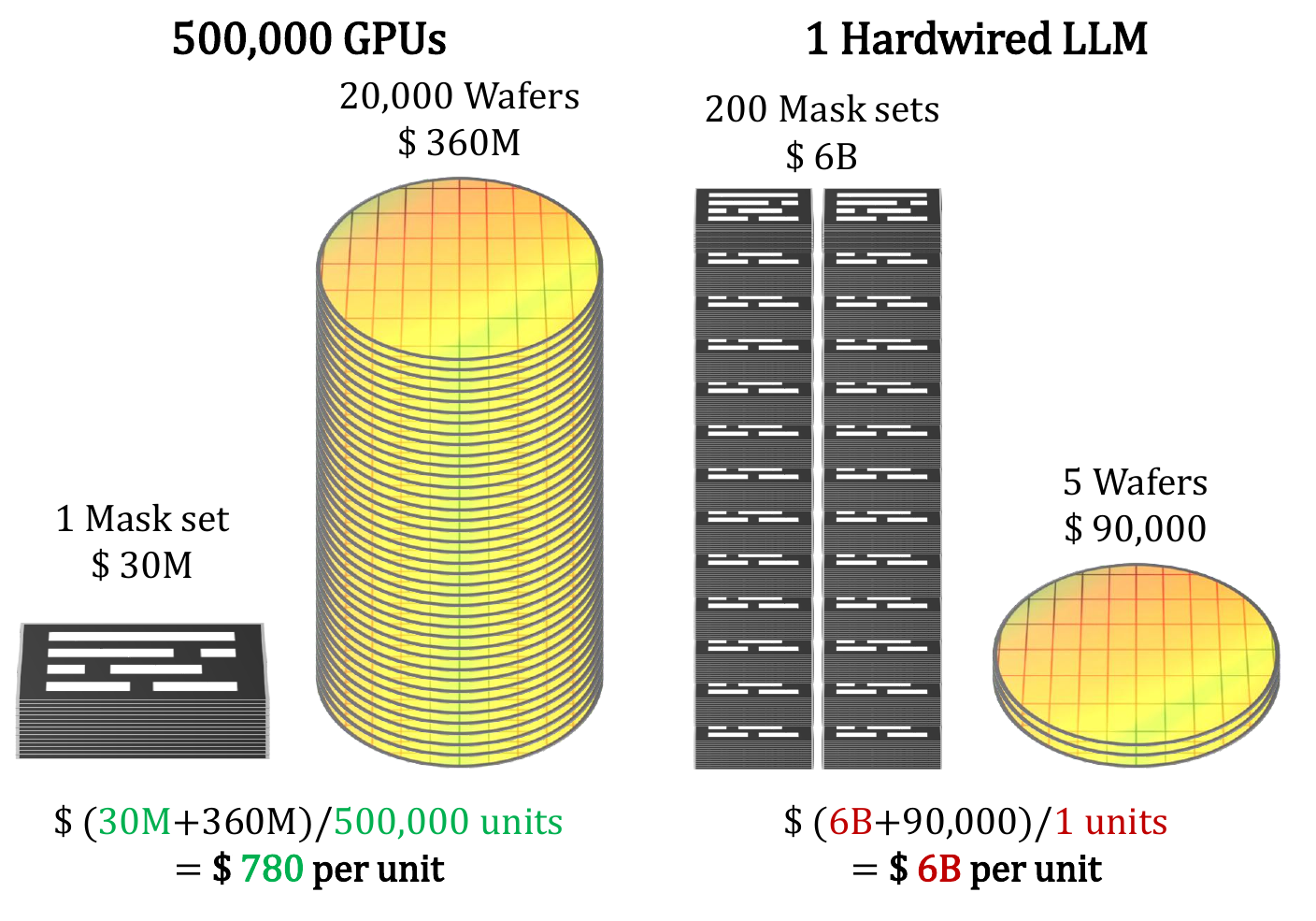}
    \caption{\revadd{\textbf{Economic challenges of hardwiring.} Considering the cost on photomasks and wafers, the cost on photomasks was amortized by the mass production of GPUs. Hardwiring an LLM incurs too many photomasks and too low volume to amortize the NRE costs.}}
    \label{fig:economics}
\end{figure}

A glaring challenge has thwarted all ambitious attempts to hardwire \revadd{LLMs}:
\emph{LLMs are too large for the current lithographic technology}.
\revadd{The practical obstacle lies in the non-recurring engineering (NRE) cost,
primarily the cost from photolithographic masks.}

\revadd{In semiconductor technology, chips are fabricated as layered structures defined by photomasks (analogous to stencils). Front-end-of-line (FEOL) processes, which form the device cells (transistors) on the silicon wafer, typically require on the order of 30 mask layers, while back-end-of-line (BEOL) processes, which form the metal interconnect stack, require approximately 40 additional mask layers. At 5\,nm technology nodes, a full photomask set is valued at \$\,30\,M, while each processed wafer is valued at approximately \$\,18,000.}
\revadd{As shown in Figure~\ref{fig:economics}, the high NRE cost of photomasks is meant to be amortized with massive production.
For example, NVIDIA is estimated manufacturing 20,000 wafers of H100 GPU, then the amortized photomask cost would be \$\,1,500 per wafer.
However, the expected production volume of hardwired LPUs is very low.
Due to the extreme computational efficiency, tens of wafers would oversaturate current LLM service demand on the planet, which renders the expensive photomask set for few uses only.}

\revadd{A straightforward implementation would require an extraordinarily large silicon area to embed the weights, and a correspondingly large number of photomask sets.}
For example, the typical transistor density of high-density 5\,nm technology is around 138\,MTr/mm\textsuperscript{2}.
FP4 Constant MAC (CMAC) requires 200+ transistors.
This translates \textsc{gpt-oss} 120\,B into an ideal area estimation of 176,000\,mm\textsuperscript{2}\revdel{, four maximal inscribed rectangles in 300\,mm wafer} \revadd{divided into 200+ chips}.
Due to reticle size limits, this hardwired LPU \revadd{must be split across 200+ photomask sets}.
Unlike prior wafer-scale practices, these masks are heterogeneous \revadd{because each chip is carrying different parts of the model weights}.
This incurs \revtarget{calc1}\revadd{\$\,30\,M $\times$ 200 = \$\,6\,B NRE} costs on photomask making, rendering a straightforward hardwired LPU economically prohibitive.

Compounding this economic infeasibility is the fast LLM development cycles.
The \revadd{prohibitively} high NRE is even worse when considering the annual parameter updates and 3-year typical lifespan of production LLMs (e.g., GPT-4).
This multiple-orders-of-magnitude cost gap is unlikely to be bridged solely through model compression.
\revadd{Note} that the \textsc{gpt-oss} is already FP4 by default. The model size has a concrete lower bound implied by the Kolmogorov complexity of human knowledge representation.

To realize the concept of hardwired LPUs, there must be fundamental breakthroughs in on-die weight embedding methodologies.
By co-optimizing with lithographic technology factors, we can achieve multiple orders-of-magnitude reduction in photomask count,
eventually clear the path to the economically viable hardwired LPUs.

%% file: tex/design.tex
{\section{Metal-Embedding}}
\label{sec::principle}

\begin{figure*}
    \centering
    \includegraphics[width=\textwidth]{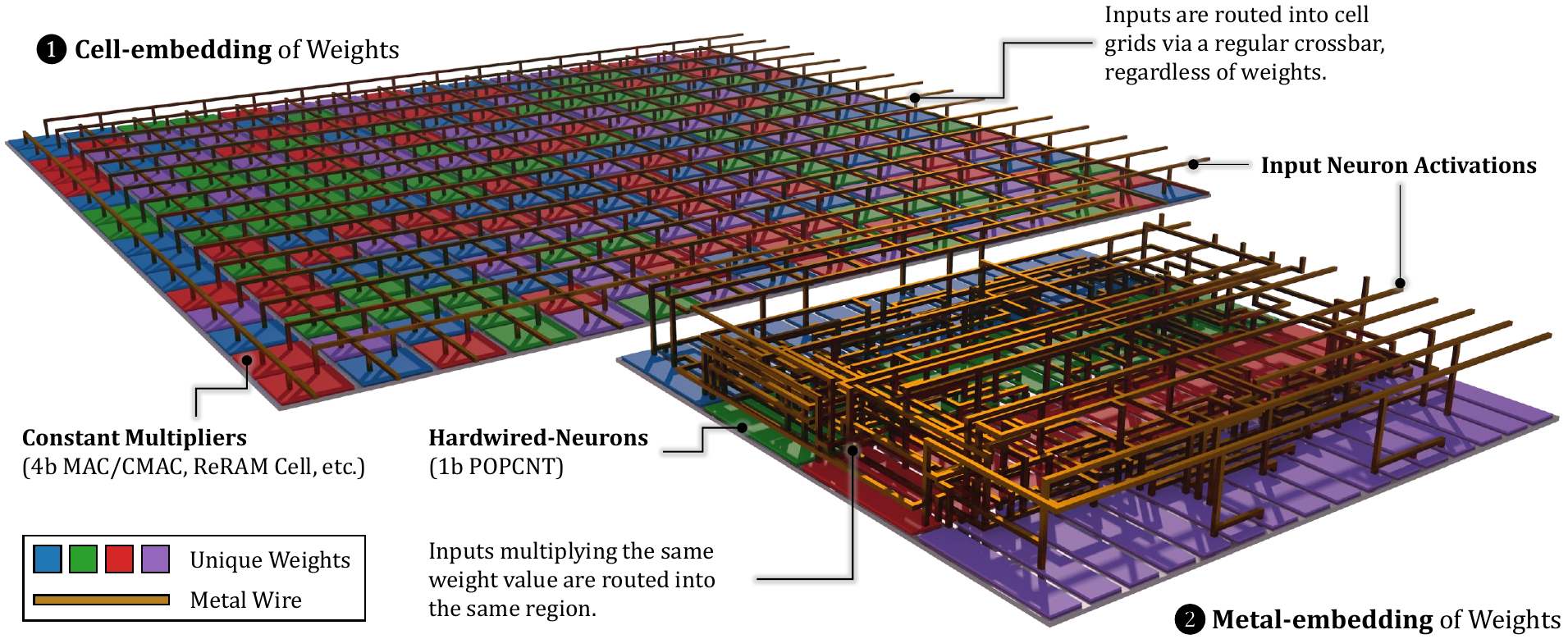}
    \caption{\textbf{Conceptual schematic of Metal-Embedding.} \ding{182} Conventional design embeds weights \revadd{in the} 2D grid of cells. Metal layers are only for physical design (P\&R). \ding{183} Metal-Embedding dramatically increases the density by leveraging the 3D topology structure of metal wires, emulating the axon-dendrite in brains.}
    \label{fig:metal-embedding}
\end{figure*}

\revdel{Currently, almost all AI chips are designed as regular grid structures (Figure~\ref{fig:metal-embedding} \ding{182}). Inputs, weights, and outputs are transmitted into/out-of the grid by a regular crossbar of metal wires. By fixing the weights, multipliers could be optimized as multiply-by-constant which is several times simpler in Boolean complexity. For example, FP4 multiply-by-constant is 42.5\,Tr on average, $\sim$6× smaller than FP4 multiplier. However, this density level is still far behind the requirement of \uname{}.}

\revdel{In current methodologies, metal layers are only considered during place-and-route (P\&R), and the topology of metal wires do not express any specific information. We see this as a waste of resources. The key insight here is that, as the topology of metal wires are three-dimensional, they potentially could embed information in a much higher density than silicon devices. As silicon devices are the area-limiting factors in the design, routing signals through complex metal wires is almost free in both area and energy, compared with logic implemented in standard cells. Please recall that biological neurons in our brains have complex topology of axon-dendrite interconnections in the first place, only then come synaptic weights.}

\revtarget{sec3}

\revadd{We address this economical challenge with the novel \emph{Metal-Embedding} (ME) methodology.
There are two key innovations in ME.
1) The \emph{Hardwired-Neuron} (HN) architecture rearranges conventional multiply-accumulate arithmetic units into accumulate-multiply-accumulate,
and lifted the embedding of weight parameters from silicon devices into metal interconnections. This enables
2) the \emph{Sea-of-Neurons} architecture -- a metal-programmable structured ASIC saving photomasks through a prefabricated array of HNs.}

{
\subsection{Hardwired-Neuron Architecture}\revtarget{sec31}
}

\revdel{We conceptualize Metal-Embedding (ME) in Figure~\ref{fig:metal-embedding} \ding{183}.
Instead of embedding weights with 2D grid of cells (silicon devices), we choose to embed weights in 3D topology of metal wires.
The die is split into Hardwired-Neurons (HNs, each HN is drawn as a $\nwarrow$\llap{$\searrow$} row in the schematic),
each HN corresponds to an output neuron activation in the model.
Each HN is divided into several regions (different coloring in the schematic), where each region represents a unique weight value.
For 4b precision models, there are $2^4=16$ unique values (4 shown in the schematic).
To multiply with a weight, the input neuron activation signal is routed into the corresponding region via metal wires.
Specifically, each input signal is multiplying with 2,880 weights (16 shown in the schematic) in \textsc{gpt-oss} 120\,B:
The model weights are solely embedded within the metal interconnection, expressed by connecting each input signal with corresponding regions. %
}

\begin{figure}
    \centering
    \includegraphics[width=\columnwidth]{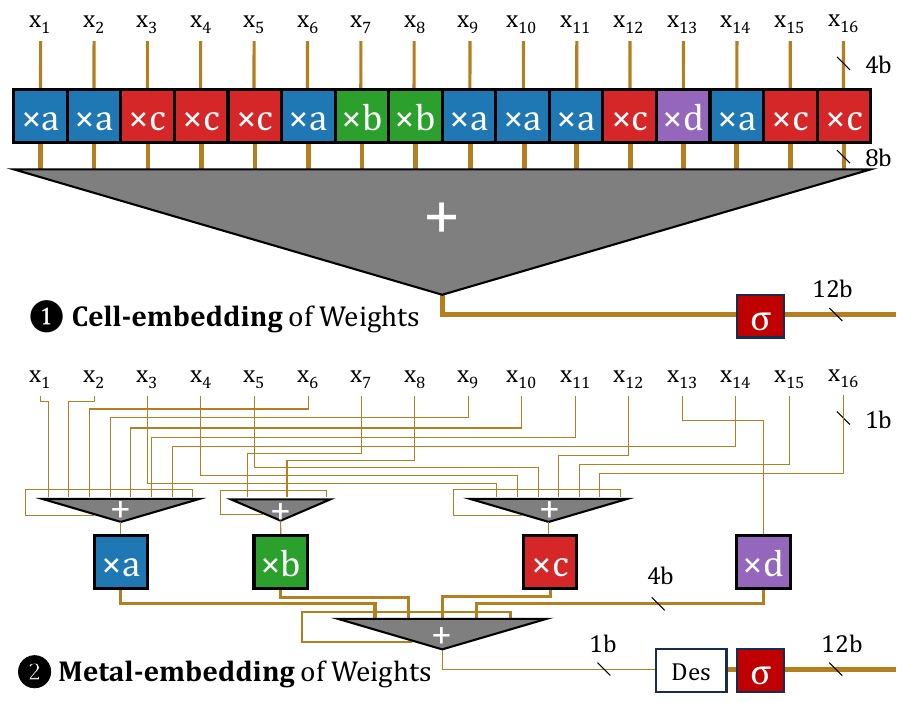}
    \caption{\textbf{Hardwired-Neuron architecture.}
    \ding{182} A conventional cell-embedding neuron contains 2,880 4b multipliers (16 shown) followed by an 8b×2,880 adder tree, where 2,880 is the hidden size in \textsc{gpt-oss} 120\,B;
    \ding{183} With ME, Hardwired-Neurons accept 1b serialized inputs (LSB-first), (1) route the inputs multiplying the same weight value to the same region, (2) perform \revadd{accumulation} (POPCNT) on these input signals, (3) perform actual multiplication with 16 multipliers (4 shown), (4) sum the results with a 4b×16 adder tree. Note how \ding{183} is significantly smaller in area than \ding{182} by reducing the number of multipliers and the strength of adders.
    }
    \label{fig:mini-arch-compare}
\end{figure}

\revdel{Then HN carries the multiplying per region, as shown in Figure~\ref{fig:mini-arch-compare} \ding{183}.
As all input signals $x_1, x_2, \dots, x_n$ in the same region are multiplying with the same constant weight value $a$,
instead of performing $ax_1 + ax_2 + \dots + ax_n$, HN performs $a\left(x_1 + x_2 + \dots + x_n\right)$ by applying the distributive law.
Furthermore, if input signals $x_1, x_2, \dots, x_n$ are in binary format, they could be serialized from the least-significant bit (LSB) to the most-significant bit (MSB) to further simplify the circuit.
After serialization, the arithmetic per region is reduced to a population count (POPCNT) followed by a constant-multiplication.
The partial results from regions are then summed with a now much smaller adder tree.}

\revdel{Specifically, a conventional cell-embedding neuron (Figure~\ref{fig:mini-arch-compare} \ding{182}) requires 2,880 4b×4b multiply-by-constant, followed by an 8b×2,880 adder tree.
A ME HN (Figure~\ref{fig:mini-arch-compare} \ding{183}) reduced the arithmetic strength to (2,880 in total)×1b POPCNT, followed by 16 4b×1b multiply-by-constant and a 4b×16 adder tree.
By removing the weight embedding from area-limiting silicon layers, ME HN reduced the area by an order-of-magnitude,
which we will demonstrate with post-layout evaluations.}

\revadd{We demonstrate the step-by-step evolution from the straightforward FP4 multiply-and-accumulate units to Hardwired-Neurons (HN). Several key arithmetic techniques are applied to minimize the required silicon area.}

\paragraph{\revadd{Basic: Weight Constancy}}\revtarget{wcon}
\revadd{Conventional hardwiring (the \$\,6\,B scenario) utilizes weight constancy.}
By fixing the weights, multipliers could be optimized as multiply-by-constant which is several times \revadd{lower} in Boolean complexity.
An FP4 multiply-by-constant unit is %
$\sim$6× smaller than an FP4 multiplier \revadd{as seen in GPU.
Accumulation could also benefit from the weight constancy under the help of optimizing EDA tools.}

\begin{figure}\revtarget{fig3}
    \centering
    \includegraphics[width=\columnwidth]{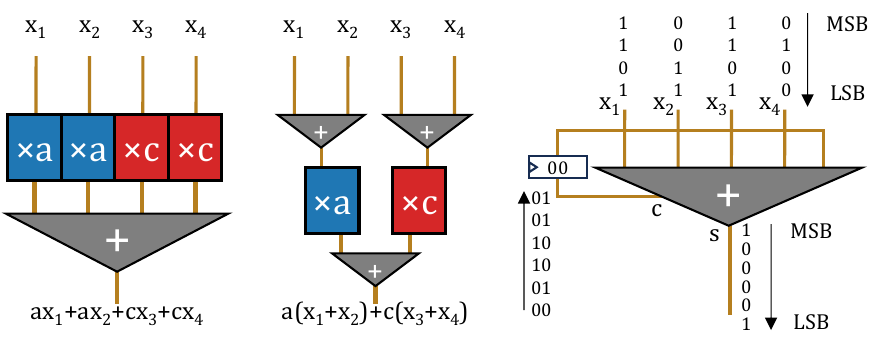}
    \caption{\revadd{\textbf{Key arithmetic techniques.} To the middle: Combining repeated multipliers via the distributive law. To the right: Using Carry Save Adders (CSA) on bit-serialized inputs to trade time for area.}}
    \label{fig:distributive}
\end{figure}

\paragraph{\revadd{Step 1: Distributive Law}}\revtarget{dlaw}
\revadd{In conventional hardwiring, FP4 weights have 16 unique values, but there are 2,880\footnote{The hidden size in \textsc{gpt-oss} 120\,B.} constant multipliers in each neuron. Most of them are repeated. By the distributive law, common multipliers could be extracted and combined. As shown in Figure~\ref{fig:distributive},} instead of performing $ax_1 + ax_2 + \dots + ax_n$ \revadd{(to the left)}, HN performs $a\left(x_1 + x_2 + \dots + x_n\right)$ \revadd{(to the middle) which saves multipliers and reduces the width of accumulation.}

\paragraph{\revadd{Step 2: Bit-serialization}}\revtarget{bits}
If input signals $x_1, x_2, \dots, x_n$ are in binary format, they could be serialized from the least-significant bit (LSB) to the most-significant bit (MSB) to further simplify the circuit.
\revadd{As shown in Figure~\ref{fig:distributive} (to the right), the single-clock-cycle accumulation could unfold into a multiple-clock-cycle tree of Carry Save Adders~\cite{hwang}, trading off speed for minimized area.}

\revadd{HN is an accumulate-multiply-accumulate unit adopting all of the above-mentioned techniques as shown in Figure~\ref{fig:mini-arch-compare}. The main result of the sophisticated combination of these techniques is that \textbf{HN lifted the embedding of weight parameters from silicon devices into metal interconnections}, as the name \emph{Metal-embedding} suggests. Conventional neurons are \emph{Cell-Embedding} (CE, Figure~\ref{fig:mini-arch-compare}\,\ding{182}), i.e., weight parameters are written into the silicon device cells composed of different constant-multipliers; HNs are \emph{Metal-Embedding} (ME, Figure~\ref{fig:mini-arch-compare}\,\ding{183}), i.e., the weight parameters are embedded as metal wires. \textbf{The silicon device cells in HN can be made parameter-independent.} }

\begin{figure}\revtarget{fig5}
    \centering
    \includegraphics[width=\columnwidth]{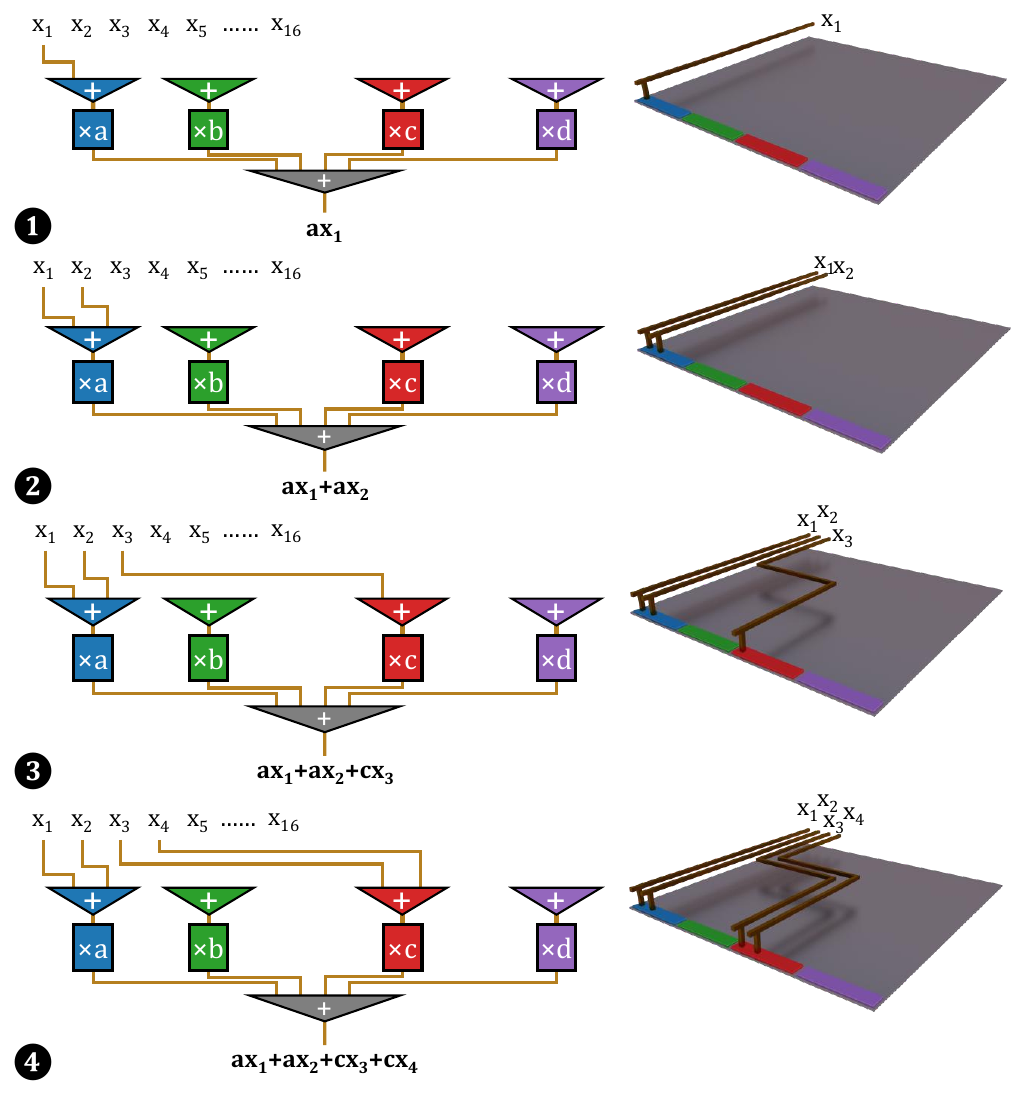}
    \caption{\revadd{\textbf{Step-by-step schematic showing how weights are physically embedded in the 3D metal wire topology.}
    HNs are accumulate-multiply-accumulate arithmetic units where each weight parameter is expressed by the source and destination of a metal wire: \ding{182} $ax_1$ by connecting from $x_1$ to the blue region; \ding{183} $ax_2$ by connecting from $x_2$ to the blue region; \ding{184} $cx_3$ by connecting from $x_3$ to the red region; \ding{185} $cx_4$ by connecting from $x_4$ to the red region.}}
    \label{fig:stepping}
\end{figure}

\revtarget{explain_emb}\revadd{Figure~\ref{fig:stepping} shows the weight embedding process through metal wires step-by-step.
For each unique value in FP4, there is an accumulator (POPCNT) in the HN. We denote the accumulators corresponding to each unique weight value with different colors (FP4 has 16 unique values, 4 shown in the schematic: blue for $a$; green for $b$; red for $c$; violet for $d$).
For each input $x_i$, the weight $w$ is embedded as a metal wire connecting input $x_i$ to the accumulator for the value $w$.
For example, as the first term is $ax_1$, a metal wire is built connecting $x_1$ to the blue accumulator (Figure~\ref{fig:stepping}\,\ding{182}). Note how the silicon devices are made parameter-independent: changing the weight from $a$ to $b$ would only change the wire destination from blue to green, while the whole accumulate-multiply-accumulate arithmetic unit is kept unchanged.
}

\revadd{To address the imbalance of weight values, the size of accumulators should be made with sufficient slackness. The accumulators could be implemented as multiple slices and be reconfigurable through metal wires. Unused ports on the accumulators are connected with zero inputs (ground).}

\revadd{Figure~\ref{fig:metal-embedding}\,\ding{183} provides an intuitive conceptual schematic of ME.}
The die is split into HNs; each HN is drawn as a $\nwarrow$\llap{$\searrow$} row in the schematic.
Each HN corresponds to an output neuron activation in the model.
Each HN is divided into several regions (different coloring in the schematic), where each region represents a unique weight value.
For 4b precision models, there are $2^4=16$ unique \revadd{weight} values (4 shown in the schematic).
To multiply with a weight, the input neuron activation signal is routed into the corresponding region via metal wires.
The model weights are solely embedded within the metal interconnection, expressed by connecting each input signal with corresponding regions.

The key insight here is that, as the metal wire topology \revadd{is} three-dimensional, they could potentially embed information in a much higher density than silicon devices.
\revadd{Current CE methodologies (Figure~\ref{fig:metal-embedding}\,\ding{182}) fail to recognize metal layers in their architecture design.}
Metal layers are only considered \revadd{in physical design (place-and-route, P\&R), thus} the topology of metal wires \revadd{does} not express any specific information. We \revadd{view} this as a waste of resources. \revadd{Since} silicon devices are the area-limiting factors in the design, routing signals through complex metal wires is \revadd{virtually} free in both area and energy, compared with logic implemented in standard cells.
\revadd{We refer to the novel architecture as \emph{Hardwired-Neurons} because of its structural similarities with biological neurons.}
Biological neurons in \revadd{the} brains have complex topology of axon-dendrite interconnections in the first place, only then come synaptic weights.
 
\revadd{When adopted alone, the HN architecture increases the density of hardwired LLM by an order of magnitude (from 200+ chips to 16 chips). But more importantly, HN concentrates all the parameter-dependent structures into metal wires, which is a prerequisite step towards introducing the \emph{Sea-of-Neurons} architecture.}

\vspace{-1em}

\subsection{Sea-of-Neurons Architecture}\revtarget{sec32}

\revadd{Up to this point, there are two common concerns to address:
\textbf{1) The NRE is still high.}}
Even with significantly reduced area of HN, 16 chips still require 16 full mask sets each valued \$\,30\,M, that is \$\,480\,M.
The total NRE still offset most economic interests.
\revadd{\textbf{2) What if the weight parameters change?}
LLM requires at least annual updates to keep competitive, and there would be unforeseen hotfixes.
Do we need another \$\,480\,M for a Hardwired LPU re-spin?}

\revadd{The key to these concerns is to \textbf{share and reuse parameter-independent photomasks}.
As the silicon devices in HN are parameter-independent, we can prefabricate HN arrays with a shared photomask set, then finalize the metal embedding wires with a few additional parameter-dependent photomask layers.
By concentrating metal embedding wires into higher level metal layers, the majority of the photomask cost can be saved.
The 16 chips could share the same photomask set for the prefabricated HN array, and the photomask set could be reused for future weight update re-spins.
}

\begin{figure}\revtarget{fig7}
    \centering
    \includegraphics[width=\columnwidth]{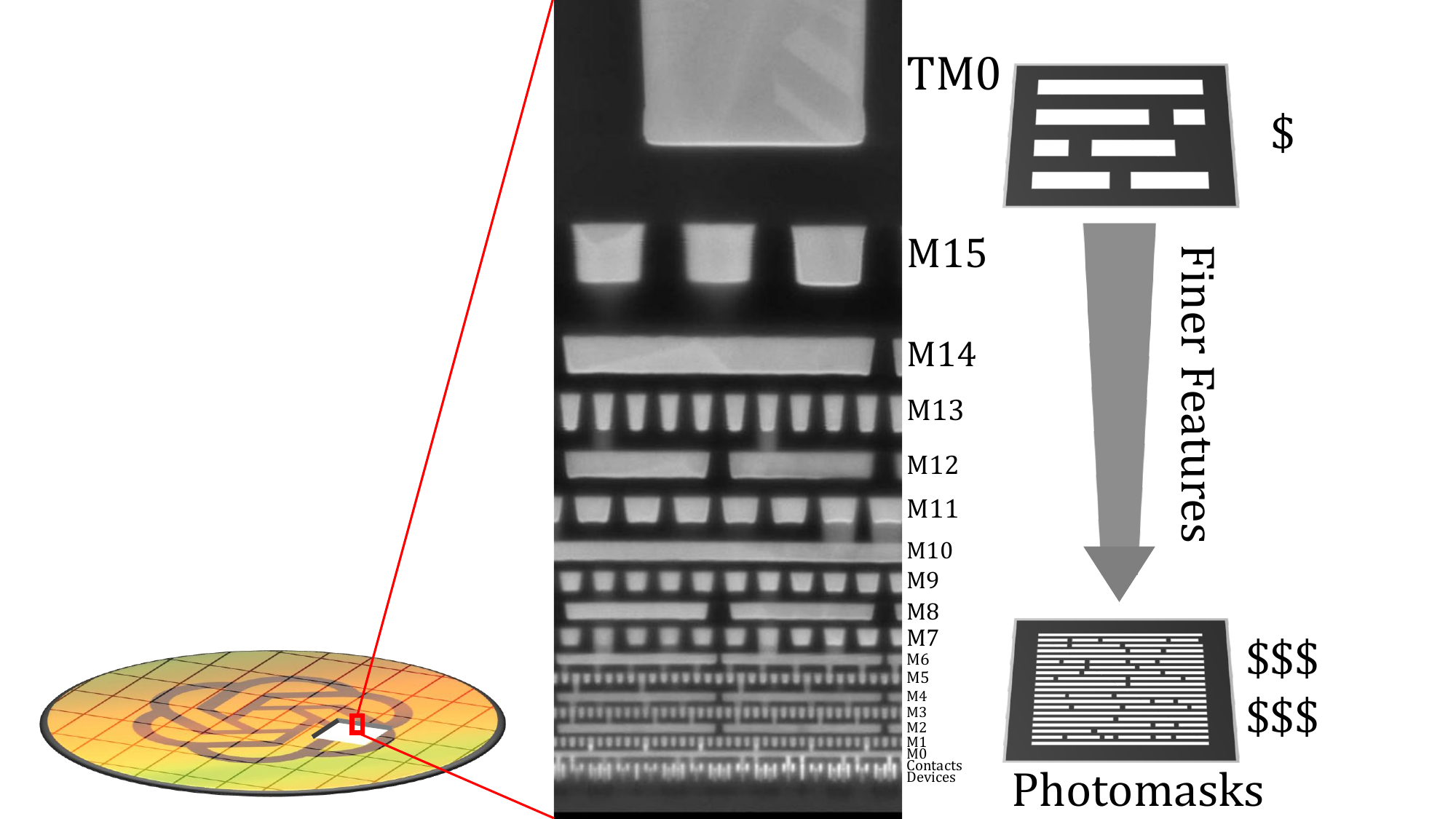}
    \caption{\revadd{\textbf{The cross section of chips~\cite{intel4}.} Geometric patterns in semiconductor chips are defined by photomasks. Silicon devices and lower metal layers have finer geometric features and thus require much more expensive photomasks to define them. }}
    \label{fig:process}
\end{figure}

\revtarget{litho_layers}\revadd{Different layers of photomask in the set are valued differently. Generally speaking, higher levels use cheaper photomasks.
As shown in Figure~\ref{fig:process}, silicon devices and lower metal layers have finer feature dimensions and requires high resolution lithographic patterning which is expensive. 
For example, metal layers at M10--M11 ($\sim$60\,nm half-pitch) require Deep Ultraviolet (DUV) single-exposure patterning (193i SE); at M4--M9 ($\sim$40\,nm half-pitch), DUV double patterning is required (typically 193i SADP, with some layers modeled as LELE); at M0--M3 ($\sim$20\,nm half-pitch), DUV quadruple patterning (193i SAQP) or Extreme Ultraviolet Lithography (EUV SE) is required.
FEOL processes making devices and contacts also require expensive EUV or DUV multiple patterning. Top metal layers including M12+ are typically reserved for power delivery networks, clock trees, and I/O peripherals. Therefore, we select M8-M11 (involving 10 layers of DUV photomasks, valued \$\,2.31\,M) as the metal-embedding layers.}

\begin{figure}\revtarget{fig8}
    \centering
    \includegraphics[width=\columnwidth]{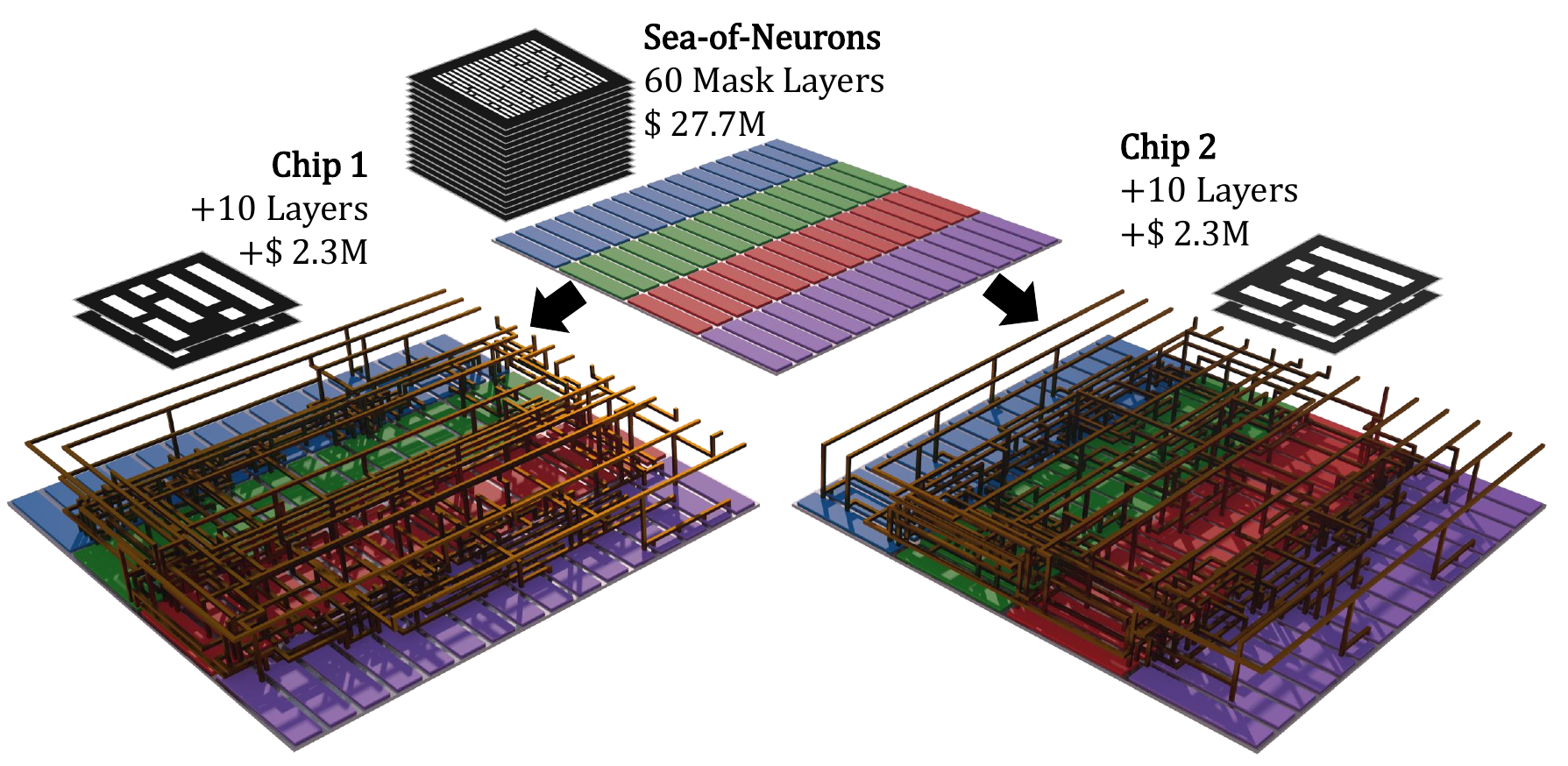}
    \includegraphics[width=\columnwidth]{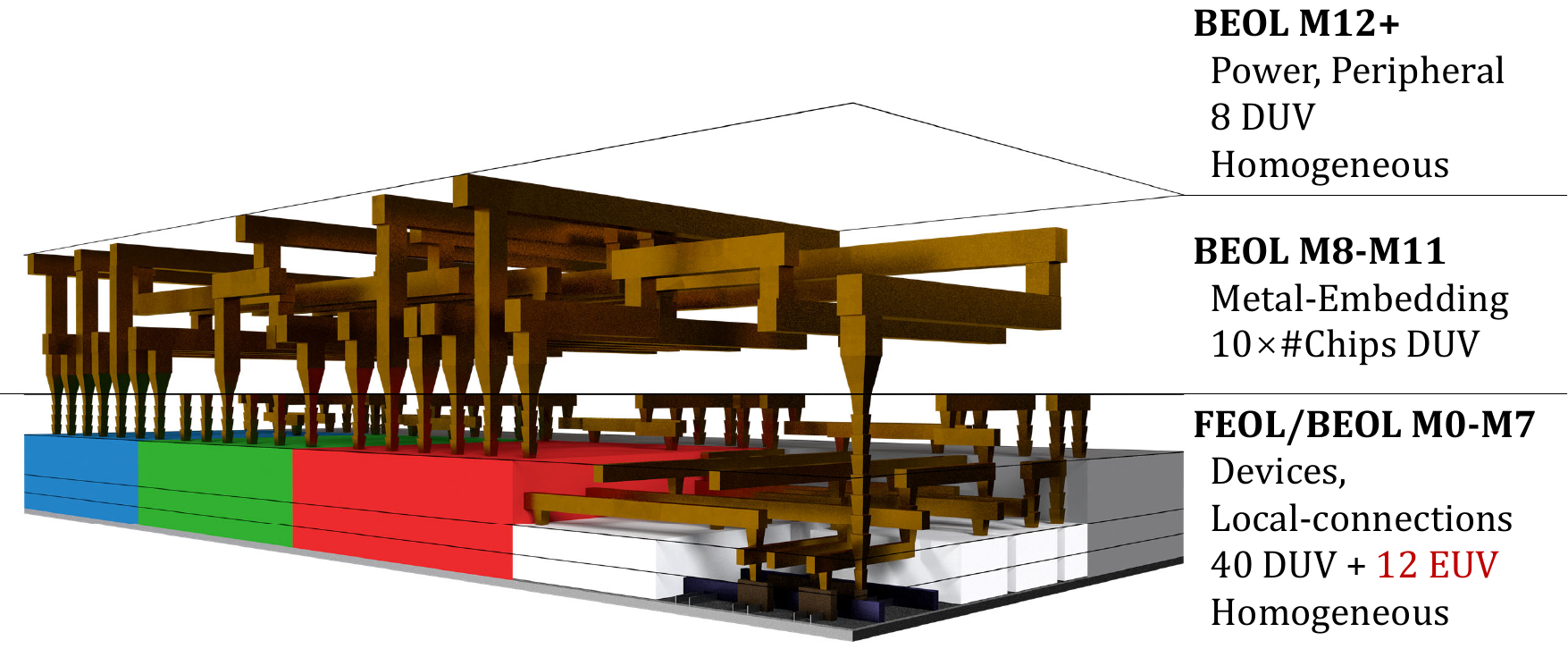}
    \caption{\revadd{\textbf{Sea-of-Neurons Architecture.} To the top: All chips share the same prefabricated HN array and 60 layers of photomask, then the embedding metalization (M8-M11) requires additional 10 layers of photomask per chip. To the bottom: The schematic from cross section with opened HN module box, showing devices and local metal wires (M0-M7) inside the HN array that are made homogeneous and mask-sharing. Note all the critical layers requiring EUV are shared.}}
    \label{fig:homomasks}
\end{figure}

\revadd{The integrated circuit design approaches to save photomask costs by semi-custom metalization over a prefabricated array of cells are known as \emph{Structured ASIC}~\cite{sasic}, and have emerged throughout history, including gate arrays in the 1970s, sea-of-gates in the 1990s~\cite{sea-of-gates}, Altera HardCopy in the 2000s~\cite{hardcopy}, and Intel eASIC N5X in 2020~\cite{easic}.
As our approach is prefabricating arrays of neurons instead of gates, we refer to it as the \emph{Sea-of-Neurons} architecture.}

\revadd{Figure~\ref{fig:homomasks} illustrates the Sea-of-Neurons architecture. Sea-of-Neurons is a metal-programmable architecture: Weights are programmed into the architecture with M8-M11 metalization over a prefabricated HN array. As 60 out of 70 mask layers are shared (including all critical layers requiring EUV), the photomask cost is significantly reduced from \$\,480\,M to \$\,65\,M \footnote{\floatingpenalty=0\relax\revtarget{calc2}\revadd{\$\,27.69\,M (the prefabricated HN array) + \$\,2.31\,M (M8-M11 metalization per-chip) $\times$ 16 (number of chips)}}. When the weight parameters change, a re-spin requires only \$\,37\,M \footnote{\floatingpenalty=0\relax\revtarget{calc3}\revadd{\$\,2.31\,M (M8-M11 metalization per-chip) $\times$ 16 (number of chips)}} as the prefabricated HN array is ready.  }

\revtarget{edatour}\revadd{The Sea-of-Neurons architecture is compatible with standard ASIC design flow and EDA tools. First, complete the P\&R of the HN array module under standard cell constraints within M0-M7.}
The layout of HN is copied to fill the major part of die area, equipped with all SoC peripherals, power grid, and clock tree.
\revadd{Next, the layout is exported to custom tools which read weight parameters and generate TCL scripts to instruct the connection of metal embedding wires.
The generated script is integrated into the overall layout within the P\&R EDA tool.
The resulting complete design is then subjected to design rule checking (DRC) and layout-versus-schematic (LVS) verification, with detected rule violations resolved through automated local repair. Finally, parasitic extraction and post-layout simulation is conducted to evaluate functional correctness and timing behavior under realistic physical effects. In our experiments, the layouts successfully completed the sign-off checks showing ample routing density margins in both M0-M7 and M8-M11.}

%% file: tex/architecture.tex
\section{Architecture}
\label{sec:architecture}

\begin{figure}[t]
  \centering
  \includegraphics[width=\columnwidth]{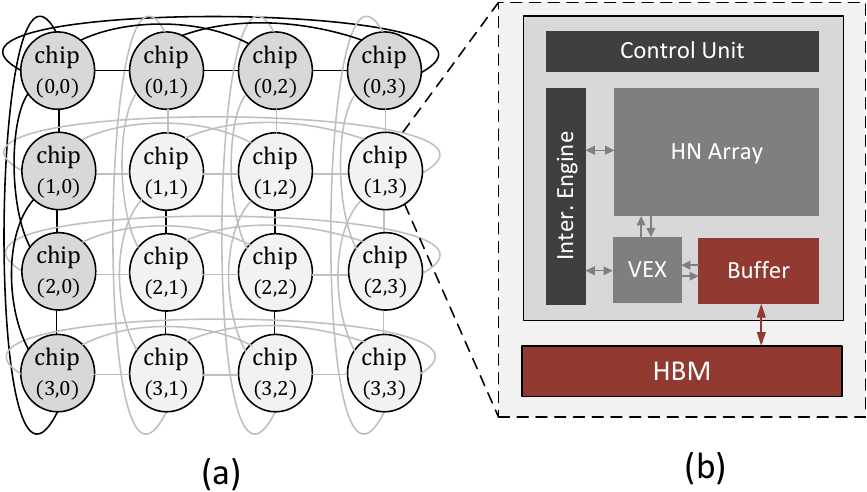}
  \caption{\textbf{The overall architecture of \uname{} system.}
  \textit{(a)} System-level architecture of \uname{}, featuring a logical $4 \times 4$ row-column fully-connected fabric to interconnect its 16 modules. %
  \textit{(b)} Architecture of a single compute module, comprising the core die and HBM. 
  }
  \label{fig:arch}
\end{figure}

In this section, we introduce the architecture of \uname{} in a top-down manner, as shown in Figure~\ref{fig:arch}, including system integration and single chip architecture.

{
\subsection{Overview}}\revtarget{sec41}

\revadd{\uname{} is a complete physical implementation of \textsc{gpt-oss} 120\,B and its computational process for inference. \uname{} system directly implements Continuous Batching on hardware to fill its pipeline. The hardware receives token IDs and generates token IDs as outputs, operating without a software stack (OS, runtime, library, compilers, frameworks). This pure hardware implementation offers two benefits: 1) It eliminates the heavy software development and maintenance cost. 2) It eliminates the software turbulence and brings more deterministic and predictable system behavior.}

\revadd{\uname{} distributes the weights across 16 chips interconnected via CXL. Besides the HN array, the chips also implement embedding dictionary lookup, Grouped Query Attention (GQA), Mixture-of-Experts (MoE) routing, Root-Mean-Square Normalization (RMSNorm), Swish-Gated Linear Unit (SwiGLU), and logit sampling. A memory subsystem is implemented for the embedding dictionary and the attention buffer (KV Cache), including SRAM and HBM. }

\subsection{System Integration}
\label{ssec::system}

\paragraph{Interconnection topology.} 

\revadd{\uname{}} system architecture is built upon a 16-module row-column fully-connected fabric. As conceptually illustrated by the logical topology in Figure~\ref{fig:arch}\textit{(a)}, this fabric establishes direct, point-to-point links from each module to all other modules within its row and, simultaneously, to all other modules within its column. This design creates a router-less\revadd{,} low-latency network for efficient collective communication patterns (e.g., All-Reduce). Each compute module is a self-contained unit, equipped with a dedicated HBM for storing the KV Cache and the embedding tables.

\paragraph{Multi-chip group mapping.}
\uname{} evenly distributes its constituent chips into multiple row- and column- groups, 
with each row and column containing 4 chips. %
This grouping strategy enables a parallel mapping of the self-attention and feed-forward network—the most computationally intensive parts of a Transformer block.
Specifically, 
\begin{enumerate}[noitemsep,topsep=0pt,parsep=0pt,partopsep=0pt,leftmargin=*]
\item For the GQA projection,
the projections for all query, key, and value heads are uniformly mapped to their respective column groups. 
\item For attention score,
 query-head\revadd{s are} all-reduced within the same column groups, while key- and value- heads are reduced to the chip-($\ell\bmod 4$), where $\ell$ denotes the sequence length. 
\item For the feed-forward network with MoE, 
all experts are uniformly distributed to all chips, and the input vector broadcasts to all chips. Specifically, each chip is responsible for 8 experts. 
\end{enumerate}

This group mapping strategy of \uname{} offers two key advantages. 
First, by distributing the GQA computation uniformly, the workload is balanced across all chips. This alleviates pressure on key computational resources (e.g., VEX units) and reduces the storage and bandwidth demands on the SRAM and HBM.
Second, the independence of the MoE experts enables fully parallel FFN computation, eliminating the need for data exchange during the projection steps.
The detailed execution process and dataflow are further elaborated in Section~\ref{sec:dataflow}.

\paragraph{Physical System Integration}
The physical implementation of \uname{} is based on established, industry-validated High-Performance Computing (HPC) integration practices.

\textbf{Packaging}: Each compute module utilizes 2.5D packaging to integrate a large monolithic die with its dedicated HBM stacks (conceptual topology is similar to the NVIDIA Blackwell platform).

\textbf{Inter-Chip Communication}: Direct point-to-point interconnects are established via the CXL 3.0 protocol (on PCIe PHY). This open standard offers low latency (<100 ns) and high bandwidth (128 GB/s per $\times$16 link), with performance approaching proprietary solutions (e.g., NVIDIA NVLink).

\textbf{Manufacturing Yield}: The modular design enables a "Known-Good-Module" strategy. Each packaged module is tested independently, thus decoupling the final system's assembly yield from the challenging manufacturing yield of the large monolithic dies.

\textbf{Thermal Management}: For high thermal density, a Direct-to-Chip Liquid Cooling (DLC) solution is employed by mounting a cold plate on each module—an approach validated in compute platforms such as the NVIDIA DGX H100.

\subsection{Single Chip Architecture}
\label{ssec::single-chip-arch}

As shown in Figure~\ref{fig:arch}\textit{(b)}, each chip in \uname{} is composed of five primary modules. 
The HN Array and VEX Unit are responsible for the LLM computation, including operations on hardwired weights, attention mechanisms, and nonlinear activation functions. The Attention Buffer serves as the on-chip KV Cache. Finally, the Control Unit manages on-chip scheduling and inter-layer pipelining for multi-batch scenarios, while the Interconnect Engine facilitates inter-chip communication.

\textbf{The HN Array} is a dedicated unit for performing computations that involve the \emph{fixed and pre-trained} weights.
As shown in Figure~\ref{fig:mini-arch-compare}.\ding{183}, we use metal embedding strategy to hardwire all the weights in the LLM onto the chip. Although the HN Array has a large area, its power consumption is remarkably low. This efficiency stems from the high sparsity of circuit activity: only 4 out of 128 experts are active at any given time in the target MoE architecture. 
For weight matrices (e.g., $W_{qkv}$) that are partitioned across multiple chips, each HN Array computes a partial sum. This result is then forwarded to the Interconnect Engine for aggregation with corresponding partial sums from other chips.

\textbf{The Vector Execution Unit (VEX)} is responsible for executing vector and matrix operations, including calculating attention scores, applying nonlinear functions (e.g., RMSNorm, SwiGLU, softmax), performing residual additions, and handling output sampling.
1) VEX adopts the FlashAttention computation flow to calculate attention scores. The hardware implementation consists of GEMV units and nonlinear operators. It fetches queries from the Interconnect Engine and reads keys/value (K/V) pairs from the on-chip Buffer. For each chip, the VEX unit is designed to process 32 cached KV-heads per cycle without stalling.
2) VEX also integrates dedicated nonlinear modules for the efficient computation of RMSNorm, SwiGLU, and softmax operations. 
Additionally, it includes a vector-aligned adder for residual connections and a specialized unit to perform multinomial sampling.

\textbf{Attention Buffer}. 
\revtarget{sec43}\revadd{The on-chip 320\,MB Attention Buffer comprises 20,000 banks, each with a 16\,KB capacity. Every bank features a 1W1R (one-write, one-read) port configuration with a 32-bit access width.}
The Attention Buffer primarily functions as a KV Cache for the chip's assigned attention groups. 
It offloads excess KV entries to HBM only when the on-chip capacity is exceeded.
This buffer also stores activation vectors for residual connections in the FFN blocks.

\textbf{Interconnect Engine and Control Unit.}
The Interconnect Engine and Control Unit on each chip jointly manage all inter-chip communication and data collectives. The communication topology is organized into row-wise and column-wise groups, each with specific supported operations:
1) For row-wise communication, the system supports a \emph{Broadcast} to distribute data (e.g., the activation vector) to all chips within the same row, and a corresponding \emph{Reduce} operation to aggregate the partial sums computed by each chip in that row.
2) For column-wise communication, the system distributes inputs to chips within a column using either a \emph{Scatter} operation, which provides each chip with a distinct portion of a vector, or a \emph{Broadcast} operation, which provides all chips with the identical vector. To collect the results, the system supports both \emph{Reduce} for aggregating partial sums and \emph{Gather} for concatenating output vectors.

%% file: tex/dataflow.tex
\section{Execution Dataflow}
\label{sec:dataflow}

In this section, we introduce the multi-chip interconnect dataflow of \uname{}, covering the model-to-chip mapping and the computing process of a Transformer block. In Section~\ref{ssec:dataflow_overview}, we provide an overview of \revadd{\uname{}} dataflow. Section~\ref{sec::pipeline} details our pipelining strategy and batching inference scheduling.
\revadd{Detailed description of the dataflow is presented in Appendix~\ref{sec:appendix}.}

\begin{figure}
    \centering
    \includegraphics[width=\columnwidth]{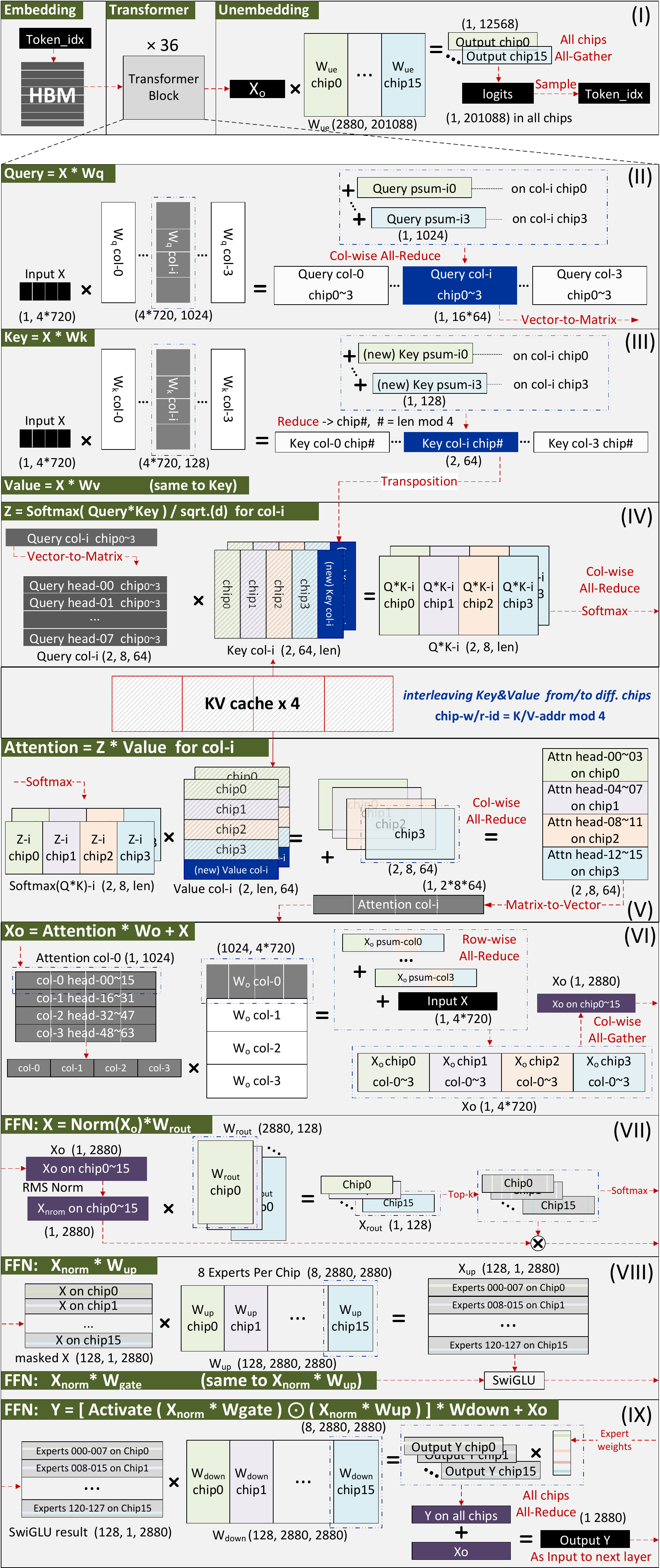}
    \caption{\textbf{Dataflow and mapping of \uname{}.}
    (\uppercase\expandafter{\romannumeral 1}) Overview of dataflow.
    (\uppercase\expandafter{\romannumeral 2}) Query projection. 
    (\uppercase\expandafter{\romannumeral 3}) Key and Value projection. 
    (\uppercase\expandafter{\romannumeral 4}\&\uppercase\expandafter{\romannumeral 5}) Attention score.
    (\uppercase\expandafter{\romannumeral 6}) Post-attention residual addition.
    (\uppercase\expandafter{\romannumeral 7}) Router in MoE. 
    (\uppercase\expandafter{\romannumeral 8}) Up- and Gate-projection. 
    (\uppercase\expandafter{\romannumeral 9}) Down-projection and residual addition.}
    \label{fig:dataflow}
\end{figure}

\begin{figure*}
    \centering
    \includegraphics[width=\textwidth]{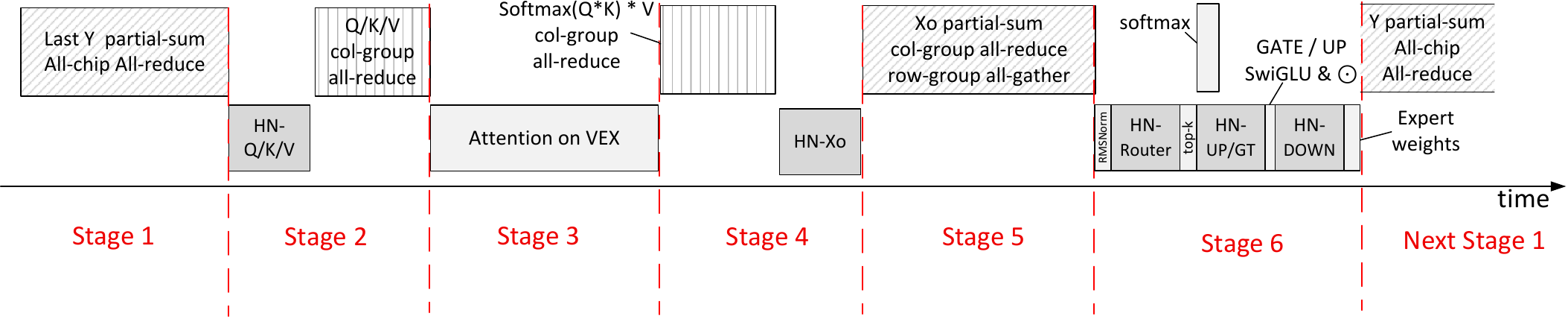}
    \caption{A six-stage pipeline partitioning diagram for \uname{} dataflow.}
    \label{fig:stage}
\end{figure*}

\subsection{Dataflow Overview}
\label{ssec:dataflow_overview}

Figure~\ref{fig:dataflow} illustrates the dataflow of \uname{}. Our design is driven by the primary goals of distributing computational load, KV cache memory access, and thermal loads, while minimizing inter-chip data communication.

Processing begins with fetching a token vector of shape $(1,2880)$ from the High-Bandwidth Memory (HBM) based on the received token index. This vector then traverses through 36 transformer blocks, where self-attention and feed-forward network (FFN) computations are performed layer by layer. 
For the multi-head attention module, we employ a hybrid weight distribution strategy: the $W_{qkv}$ weight matrix is partitioned column-wise across different chip column groups, while the $W_o$ weight matrix is partitioned row-wise across the same set of chips.
This design allows for parallel, independent computation across different chip columns for the $W_{qkv}$ operations and across different chip rows for the subsequent $W_o$ computations, thereby minimizing inter-chip data transfer.

The FFN implementation, which uses a Mixture-of-Experts (MoE) architecture, assigns eight experts to each chip, allowing for entirely independent computation with no inter-chip communication. A key exception to our partitioning strategy is the \revadd{router} weight matrix, $W_{\text{rout}}$, which is replicated across all chips. This deliberate design trade-off introduces a negligible area overhead—as the router's weights constitute only about 0.01\% of the total model weights—but eliminates the communication latency. Once all transformer blocks have been processed, the final output vector is passed through the unembedding layer, and a new token index is calculated through a sampling operation. Detailed description of the dataflow \revadd{is} presented in Appendix~\ref{sec:appendix}.

\subsection{Pipeline and Scheduling}\label{sec::pipeline}

\uname{} employs a nested pipelining strategy to maximize throughput, consisting of both inter-layer and intra-layer pipelining.
Since all layer weights are hardwired onto the metal layers, the wiring weights for each layer have their own corresponding computing resources. HNs of each layer can operate simultaneously, which facilitates the straightforward formation of a pipeline between the model layers.
Within a layer, we partition the computation into a six-stage pipeline, as shown in Figure~\ref{fig:stage}.
Consequently, \uname{} can process up to \revadd{(6 $\times$ \#layer) requests} simultaneously at peak. For the 36-layers LLM, the maximum batch size can theoretically reach 216.

\uname{} employs the batching strategy similar to Continuous Batching~\cite{yu2022orca}. 
During the \textbf{prefill} phase, there are no dependencies between the input tokens of a sequence. This independence allows for massively parallel processing, with tokens flowing through the pipeline stage-by-stage. Consequently, \uname{} can process up to 216 tokens concurrently during prefill.
Conversely, the \textbf{decode} phase is auto-regressive, meaning the generation of each new token is dependent on the completion of the previous one. However, since different sequences are independent, \uname{} can still process up to 216 sequences simultaneously.
In summary, \uname{} supports a maximum batch size of 216 sequences. By leveraging Continuous Batching, the system dynamically schedules new sequences into the batch as soon as slots are freed by completed ones, thereby ensuring high throughput.

%% file: tex/methodology.tex
\section{Methodology}
\label{sec::methodology}
This section details the methodology used to evaluate proposed \uname{} architecture. We describe our hardware evaluation flow, the system-level modeling for multi-chip design, the model used for performance assessment, and the configurations of all baseline systems.

\subsection{Hardware and System-Level Evaluation}
\label{ssec::hardware_methodology}
\paragraph{Hardware Implementation}
We implemented the core components of \uname{} architecture, including HN Array, Control Unit, VEX, Interconnection Engine and on-chip Attention Buffer in RTL with Verilog, and \revadd{verified} the correctness of the RTL design \revadd{using} extensive test cases. We followed a standard ASIC design flow to obtain physical characteristics. The design was synthesized using Synopsys Design Compiler and placed-and-routed using Synopsys IC Compiler, on 5\,nm technology. Power consumption was analyzed by PrimeTime PX using workload-derived switching activity (SAIF file) to accurately model both static and dynamic power. On-chip SRAMs were generated and analyzed using Memory Compiler on the same technology node. %

\paragraph{Multi-chip System Modeling}
Our proposed \uname{} is a $4\times4$ multi-chip system interconnected via the CXL 3.0 protocol. We evaluated the inter-chip communication latency and power using CNSim~\cite{Evaluating24Ma}, a state-of-the-art open-source analysis framework for multi-chip systems. This framework allows for detailed modeling of the network on package topology, accounting for physical layer (PHY) latency, protocol overhead, and physical routing delays in our design. We also built a cycle-level simulator for single-chip performance evaluation.

\subsection{Model}

We selected the OpenAI \textsc{gpt-oss} 120\,B model for system-level evaluation. It is a state-of-the-art open-source MoE large language model built on Llama-style architecture. We used 4-bit quantized version of the model and hardwiring the weights in HN Array. \revadd{\uname{}} implementation of the model follows partitioning method, dataflow and mapping strategies detailed in Section~\ref{sec:architecture} and Section~\ref{sec:dataflow}.

\subsection{Baseline Configurations}
We conducted two sets of experiments to comprehensively evaluate our architecture against relevant baselines.

\paragraph{Embedding Methodology Comparison}
This benchmark compares the performance of a single matrix-vector multiplication: $1 \times 1024$ input vector with a $1024\times128$ FP4 weight matrix (typical dimension in an LLM attention block) under various embedding methodologies.
We compare three designs at 5\,nm technology: \textbf{MAC Array (MA)}, a 64\,KB SRAM companioned with a conventional computing array of 1024 MACs, \textbf{Cell-Embedding (CE)} and \textbf{Metal-Embedding (ME)} as illustrated in Figure~\ref{fig:mini-arch-compare}. Regarding area, we compare CE and ME with the 64\,KB SRAM only, excluding the arbitrarily-sized computing array. 

\paragraph{System-Level Performance Comparison}
This experiment compares our full \uname{} architecture against leading commercial systems running the same \textsc{gpt-oss} 120\,B model with a 2K token length, with hyperparameters for each system individually tuned to achieve its optimal throughput.
\begin{enumerate}[noitemsep,topsep=0pt,parsep=0pt,partopsep=0pt,leftmargin=*]
    \item \textbf{NVIDIA H100}: We conducted direct measurements on a server equipped with H100 (80\,GB memory, 3.35\,TB/s bandwidth) GPU. The model was deployed via TensorRT-LLM, and the reported figures are averaged over multiple runs.
    \item \textbf{Cerebras WSE-3}: 
    The throughput was empirically measured through publicly accessible Cerebras cloud service~\cite{cerebras_inference} running the \textsc{gpt-oss} 120\,B model. %
    As power measurement on cloud is not practical, we adopted the system power figures reported in~\cite{kundu2025comparison} instead.
    
    \item \textbf{\uname{}}: %
    \revdel{The full system modeled as previous described.} %
    \revtarget{sec63}\revadd{We utilize post-PnR simulations capturing physical layout parasitics and wire delays. This approach provides high-fidelity performance projection, as \uname{} operates on a deterministic Token-In-Token-Out execution model free from software-stack variability.}
    
\end{enumerate}

%% file: tex/experiments.tex
\section{Evaluation} %
\label{sec::experiments}
\begin{figure}
    \centering
    \includegraphics[width=\columnwidth]{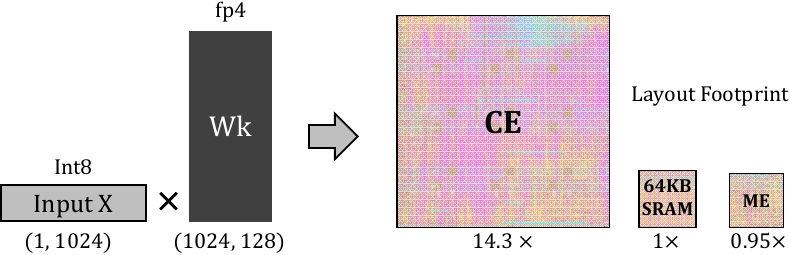}
    \caption{Area Comparison.}
    \label{fig:layout_compare}
    \vspace{-1em}
\end{figure}

\begin{figure}
    \centering
    \includegraphics[width=\columnwidth]{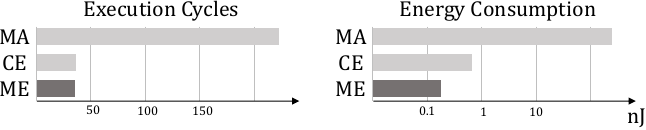}
    \caption{Time and Energy Comparison.}
    \label{fig:PPA_compare}
    \vspace{-1em}
\end{figure}

\subsection{Layout Characteristics}

\begin{table}[h!]
  \centering
  \begin{threeparttable}
  \caption{Single Chip Hardware Characteristics}
  \label{table:breakdown}
  \footnotesize
\begin{tabular*}{0.95\columnwidth}{@{\extracolsep{\fill}} l c c c c} 
   \toprule
   \        & {\textbf{Area (mm\textsuperscript{2})}} & {\textbf{\%}} & {\textbf{Power (W)}} & {\textbf{\%}} \\
   \midrule
    HN Array           & 573.16 & 69.3 & 76.92 & 24.94 \\
    VEX                & 27.87 & 3.4 & 33.09 & 10.73\\
    Control Unit       & 0.02 & 0.0 & <0.01 & 0.0\\
    Attention Buffer   & 136.11 & 16.5 & 85.73 & 27.80\\
    Interconnect Engine & 37.92 & 4.6 & 49.65 & 16.10\\
    HBM PHY   & 52 & 6.3 & 63 & 20.43\\
    \midrule
    \textbf{Total}  & \textbf{827.08} & 100.0 & \textbf{308.39} & 100.0\\
   \bottomrule
\end{tabular*}

    \end{threeparttable}
\end{table}

\revtarget{feas}\revadd{To validate the physical feasibility of \uname{}, we conducted a sign-off-grade implementation flow across representative PVT corners. 
The design achieves timing closure at 1.0\,GHz under worst-case conditions (SSG, 0.675\,V, 125\,°C), ensuring robust operation under extreme process variations and voltage drops. 
The design achieves a congestion-free layout with zero overflow. 
\revtarget{routing_density}The routing density on ME layers (M8--M11) remains below 70\% (lower than typical accelerators), validating the feasibility of ME strategy.
Signal integrity is confirmed by parasitic extraction (avg. $R=164\,\Omega$, $C=7.8$\,fF), showing manageable coupling effects. 
Thermal analysis confirms that the power density (avg. 0.3\,W/mm\textsuperscript{2}, peak 1.4\,W/mm\textsuperscript{2}) is well within the cooling limits of 2.5D packaging. 
Finally, the layout is DRC/LVS clean, and yield estimation based on Murphy’s model (defect rate 0.11/cm\textsuperscript{2}) confirms the manufacturability of the design.}

Table ~\ref{table:breakdown} presents the area and power breakdown of a single chip in \revadd{\uname{}}. The HN Array and the Attention Buffer are the dominant components in terms of both area and power. The chip occupies a total area of 827.08\,mm\textsuperscript{2} and has a power consumption of 308.39\,W.

\revtarget{mbw}\revadd{The Attention Buffer sustains 80\,TB/s bandwidth and 3-cycle latency under worst-case PVT conditions, confirming sufficient performance margins.}

The power density of the HN array is significantly lower than other components due to the sparse circuit activity induced by MoE. Specifically, only 4 out of 128 experts are activated at a time.

\subsection{Embedding Methodology Comparison}
Figure~\ref{fig:layout_compare} presents the post-layout area comparison using the SRAM in MA as a base unit.
The area of CE/SRAM(MA)/ME is 14.3$\times$/1$\times$/0.95$\times$, respectively, validating the claimed density advantage of ME.

The performance and energy consumption results are illustrated in Figure~\ref{fig:PPA_compare}. Both the ME and CE designs demonstrate a dramatic reduction in execution cycles compared to the MA by fully parallelizing the computation. Constrained by the need to fetch weights from SRAM and its limited multiplier array, \revadd{the MA} requires significantly more cycles to complete the same task. 
The energy reduction of ME is also significant. It consumes the least energy by completely eliminating memory access. While the CE also eliminates power from SRAM access, its massive area leads to substantial leakage power, making it less energy-efficient than ME. The energy consumption of MA is mainly driven by  repeated, power-intensive accesses to its SRAM.

In summary, the experimental results demonstrate the comprehensive PPA superiority of the ME design at the operator level. This validates the effectiveness of ME as the fundamental building block for LLM accelerators.

\subsection{System-Level Performance Comparison}

\begin{table}%
    \centering
    \begin{threeparttable}
        \caption{System-Level Performance and Efficiency Comparison for \textsc{gpt-oss} 120\,B Inference}
        \label{tab:system_level_comparison}
        \fontsize{8}{9.5}\selectfont
        \begin{tabular}{l r r r}
            \toprule
            \textbf{Metric} & \textbf{\uname{}} & \textbf{H100} & \textbf{WSE-3}\tnote{a} \\
            \midrule
            \multicolumn{4}{l}{\textit{Core Performance}} \\
            \quad Throughput (tokens/s) & 249,960 & 45 & 2,940 \\       %
            \midrule
            \multicolumn{4}{l}{\textit{Physical Characteristics}} \\
            \quad Technology Node & 5\,nm & 5\,nm & 5\,nm \\
            \quad Total Silicon Area (\si{mm^2}) & 13,232 & 814 & 46,225 \\
            \quad System Footprint (Rack Units) & 4\,U & 1\,U & 16\,U \\
            \midrule
            \multicolumn{4}{l}{\textit{Power \& Efficiency}} \\
            \quad Total System Power (\si{\kilo\watt}) & 6.9 & 1.3 & 23.0 \\
            \quad \textbf{Energy Efficiency (tokens/kJ)} & \textbf{36,226} & \textbf{34.6} & \textbf{127.8} \\
            \quad \textbf{Area Efficiency (tokens/$($s$\cdot$\si{mm^2}$)$)} & \textbf{18.89} & \textbf{0.055} & \textbf{0.064} \\
            \bottomrule
        \end{tabular}
        
        \begin{tablenotes}
            \item[a] \small WSE-3 data is obtained from published reports~\cite{kundu2025comparison,cerebras_power,cerebras_memoryX,cerebras_system} and calibrated against performance on its public cloud service~\cite{cerebras_inference}.
        \end{tablenotes}

    \end{threeparttable}
\end{table}

This section provides a comparison of the system-level performance and efficiency of \revadd{\uname{}}, NVIDIA H100, and Cerebras WSE-3 on \textsc{gpt-oss} 120\,B model, as detailed in Table \ref{tab:system_level_comparison}. \revadd{\uname{}} demonstrates orders-of-magnitude advantages in both throughput and energy efficiency, achieving up to $5{,}555\times$ and $85\times$ throughput, and $1{,}047\times$ and $283\times$ energy efficiency, respectively. \revadd{\uname{}}'s superior performance and efficiency stem from a fundamental architectural redesign that diverges from conventional systems.

First, \uname{} physically hardwires model weights into the compute fabric. This creates massive, fine-grained parallelism and inherently supports ultra-high throughput inference.

Second, as a direct consequence, this design completely eliminates the need to access weights from the memory hierarchy (e.g., SRAM, DRAM), thus avoiding the immense energy cost of memory access.

Finally, \revadd{\uname{}} operates on a highly optimized, model-specific dataflow, workload partitioning, and pipelining strategy, which contrasts with the instruction-driven paradigm of GPUs. This eliminates the significant overhead from control unit such as instruction decoding, scheduling, and control flow. It ensures that nearly all time, power, and area are dedicated to effective computation.

\subsection{Execution Time Analysis}

\begin{figure}\revtarget{fig14}
    \centering
    \includegraphics[width=\columnwidth]{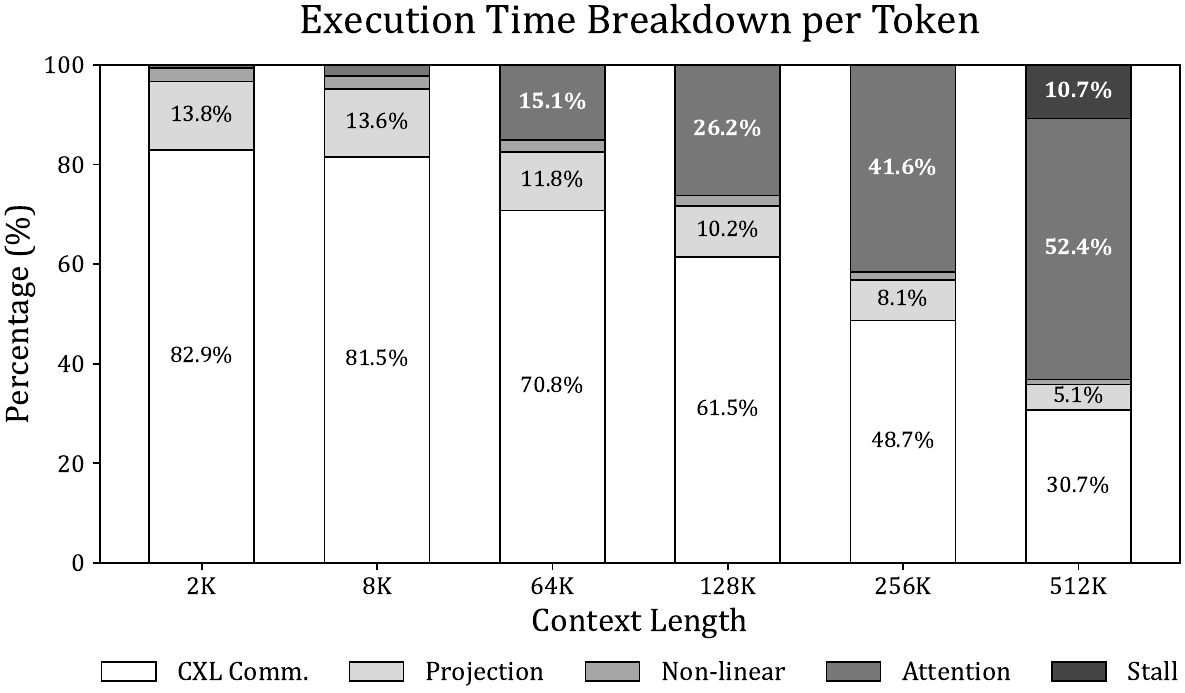}
    \caption{\revadd{\textbf{Execution time breakdown across varying context lengths.} The total execution time is decomposed into inter-chip CXL communication, projection, non-linear operations, attention computation and memory access stalls.}}
    \label{fig:time-break}
\end{figure}

\revtarget{sec74}\revadd{Figure~\ref{fig:time-break} presents the execution time breakdown across varying context lengths. Memory access latency is effectively hidden by the double-buffering mechanism: stalls remain negligible up to 256K tokens, and reach 10.7\% at an extreme context length of 512K, where KV cache is loaded from off-chip HBM. In terms of breakdown, the highly optimized computing components expose inter-chip communication as the dominant factor at shorter lengths, while attention computation becomes dominant as the sequence length increases.}

\revtarget{sec75}\subsection{\revadd{Economic Analysis} and Carbon Footprint}\label{ssec:economy}

\begin{table*}[t!]\revtarget{tab3}
    \centering
    \caption{Total Cost of Ownership (TCO) Analysis for LLM Inference over a 3-Year Lifecycle.}
    \label{tab:tco_analysis_detailed}
    \footnotesize
    \begin{threeparttable}
    {\begin{tabular*}{\textwidth}{@{\extracolsep{\fill}} l cccc }
        \toprule
        \multirow{2}{*}{\textbf{Parameter Category}} &
        \multicolumn{2}{c}{\textbf{Low Volume}} &
        \multicolumn{2}{c}{\textbf{High Volume}} \\
         & \textbf{HNLPU} & \textbf{H100} &
           \textbf{HNLPU} & \textbf{H100} \\
        \midrule

        \textit{System Configuration \& Power} \\
        \quad Number of Systems / GPUs\textsuperscript{1} &
            1 & 2,000 &
            50 & 100,000 \\
        \quad Total Datacenter Power (\si{\mega\watt})\textsuperscript{2} &
            0.010 & 3.64 &
            0.483 & 182 \\
        \midrule

        \textit{Capital Expenditure (CapEx)} \\
        \quad Node Price\textsuperscript{3} &
            \$\,59.25\,M $\sim$ 123.3\,M & \$\,79.99\,M &
            \$\,62.83\,M $\sim$ 129.9\,M & \$\,4,000\,M \\
        \quad Data Center Infrastructure\textsuperscript{4} &  %
         \$\,0.2100\,M & \$\,54.93\,M & \$\,10.30\,M & \$\,2,747\,M  \\     
        \quad \textbf{Total Initial CapEx} &
            \textbf{\$\,59.46\,M $\sim$ 123.5\,M} & \textbf{\$\,134.9\,M} &
            \textbf{\$\,73.13\,M $\sim$ 140.2\,M} & \textbf{\$\,6,747\,M} \\
        \quad Update Re-spin Cost\textsuperscript{5} &
            \$\,18.53\,M $\sim$ 37.06\,M & \$\,0.00 & 
            \$\,22.11\,M $\sim$ 43.68\,M & \$\,0.00 \\
        \midrule

        \textit{3-Year Operational Expenditure (OpEx)} \\
        \quad Electricity Cost\textsuperscript{6} &
            \$\,0.0250\,M & \$\,9.088\,M &
            \$\,1.206\,M  & \$\,454.4\,M \\
        \quad Maintenance \& Support\textsuperscript{7} & 
        \$\,0.0730\,M $\sim$ 0.1353\,M & \$\,47.24\,M & 
        \$\,0.3650\,M $\sim$ 0.6765\,M & \$\,2,362\,M \\
        \midrule

        \textit{3-Year Total Cost of Ownership (TCO)} \\
        \quad Static Model (No Updates) &
            \$\,59.56\,M $\sim$ 123.7\,M & \$\,191.2\,M &
            \$\,74.70\,M $\sim$ 142.1\,M & \$\,9,563\,M \\
        \quad Dynamic Model (Annual Updates) &
            \$\,96.62\,M $\sim$ 197.8\,M & \$\,191.2\,M &
            \$\,118.9\,M $\sim$ 229.4\,M & \$\,9,563\,M \\
        \midrule
        \textit{Sustainable AI Support} \\
        \quad Total Emissions (tCO$_2$e) (Static\,/\,Dynamic)\textsuperscript{8} & 102.0 / 106.0 & 36,600 & 4,924 / 5,124 & 1,830,000 \\
        \bottomrule
    \end{tabular*}
    \begin{tablenotes}
    \item All figures are rounded to four significant figures. Appendix~\ref{sec:appendix2} presents the detailed assumptions and source references.
    \end{tablenotes}}
    \end{threeparttable}
\end{table*}

\revadd{
We present a comprehensive Total Cost of Ownership (TCO) analysis over a three-year lifecycle in Table~\ref{tab:tco_analysis_detailed}. We compare \uname{} against an equivalently provisioned NVIDIA H100 GPU cluster delivering comparable inference throughput. 
We consider two representative deployment volumes: a low-volume deployment corresponding to a single \uname{} node, and a high-volume scenario corresponding to an OpenAI-scale deployment~\cite{openai_devday, openai_nvidia_systems_partnership_2024}. We provide both optimistic and pessimistic estimates to account for the sensitivity of key assumptions.
For detailed assumptions and source references, please refer to Appendix~\ref{sec:appendix2}.
}

\revadd{
In the low-volume scenario, \uname{} reduces the initial capital expenditure (CapEx) by 8.5--55.9\% and reduces operational expenditure (OpEx) by a factor of  351.4--574.8$\times$. 
This OpEx advantage stems from the significantly reduced physical footprint and power consumption.
Over a three-year lifecycle, even though \uname{} incurs two annual update re-spins, the TCO remains lower than, or breaks even with, that of an H100 cluster delivering equivalent throughput.
For high-volume deployments, \uname{} reduces the initial CapEx, OpEx, and TCO by factors of 48.1--92.3$\times$, 1,496--1,793$\times$, and 41.7--80.4$\times$, respectively. This increased advantage stems from amortizing the NRE costs over multiple sets of \uname{}.
}

\revadd{
Finally, we estimate the three-year equivalent carbon dioxide emissions. The carbon footprint of \uname{} is 357.2$\times$ and 371.7$\times$ lower than that of the H100 cluster, with and without annual update re-spins respectively. This is attributed to significant reductions in both hardware manufacturing (embodied carbon) and power consumption (operational carbon).
}

%% file: tex/conclusion.tex
\section{Discussion}

\revadd{\textbullet{} \textbf{Inference Volume.} Section~\ref{ssec:economy} analyzed low (single node volume) and high (50 nodes, OpenAI-scale) volumes. These volume settings are based on existing businesses.} We anticipate that the unprecedented performance of \uname{} will unlock novel LLM application scenarios that were \revadd{previously infeasible}, thereby stimulating further growth in inference volume. \revadd{As production volume increases,} NRE \revadd{costs are} further amortized, amplifying the cost advantages.

\revadd{\textbullet{} \textbf{Field-programmable vs Metal-programmable.}\revtarget{dynr} 1) As the Sea-of-Neurons architecture reduces the weight update re-spin cost to a minor fraction of the TCO, we expect no strong interest towards field-programmable architecture.
2) Introducing area overhead (more chips) to implement dynamic routing would put even more pressure on the dominant bottleneck of the multi-chip interconnection (Figure~\ref{fig:time-break}). Advanced interconnection technology (e.g., wafer-scale integration) would put both \uname{} and field-programmable LPU in a stronger position.}

\revadd{\textbullet{} \textbf{Scalability.}} We estimate the initial NRE cost on making \uname{} chips for various LLMs other than \textsc{gpt-oss} in Table~\ref{tab:cost_comparison}.
Results suggest that a wide range of model sizes can be deployed within an acceptable budget.

\begin{table}[h!]
    \centering
    \caption{Chip NRE prices on various models.}
    \label{tab:cost_comparison}
    \resizebox{\columnwidth}{!}{%
    \begin{tabular}{lccccc}
    \toprule
    & Kimi-K2~\cite{team2025kimi} & DeepSeek-V3~\cite{liu2024deepseek}  & QwQ~\cite{qwq} & Llama-3~\cite{llama3} \\
    \midrule
    Param. \# & 1\,T & 671\,B  & 32\,B & 8\,B \\
    Price/M\$ & 462 & 353  & 69 & 38 \\
    \bottomrule
    \end{tabular}%
    }
\end{table}

\revadd{\textbullet{} \textbf{Yield and Fault Tolerance.} Unlike mass-produced processors, yield is a secondary factor to \uname{}. Assumption of 1\% yield implies producing $\sim$50$\times$ more wafers than calculated in Table~\ref{tab:tco_analysis_detailed}. These wafers cost \$\,0.5\,M/\$\,22\,M in low/high volume CapEx, which are marginal compared to the TCO.}

\revtarget{blue_green}\revadd{\textbullet{} \textbf{Model Updates.} \uname{} updates are performance steps. There is no task that GPT-5.2 can handle but GPT-5.1 cannot attempt, just as the release of B100 did not render H100 obsolete.
The "blue-green" deployment model can be adopted for seamless updates:
When a model update is validated on GPU testbeds, new "green" \uname{} can be manufactured while the "blue" \uname{} continue serving traffic.
Estimated turnaround time is 6–8 weeks. The cost is comparable with regular processor re-spins thanks to the Sea-of-Neurons.}

\revadd{\textbullet{} \textbf{Future works.}} 
(1) \emph{Enhanced Flexibility on Sea-of-Neurons}, enabling hyper-parameter updates with annual re-spin by programmable dataflow; 
(2) \emph{Automated Design and Test}, including an automated Hardwired-Neuron Compiler for shortening the delay in the design flow; 
(3) \emph{Extended Application Scenarios}, implementing conditional decoding (programmable sampling algorithms), and support of use cases other than generation (sequence scoring, text-embedding, etc.);
\revadd{(4) \emph{LoRA for Post-Deployment Updates}, adding $\sim$1\% field-programmable HNs at side-channel to accommodate dynamic weights.}
However, we \revadd{foresee} no \revadd{significant} technical obstacles \revadd{to} implementing these features on \revadd{the} \uname{}.

\section{Related Works}
This work is mostly inspired by the current \revadd{progress} in large language models.
The success of GPT-3 has proven that LLMs are few-shot learners~\cite{gpt3}.
DeepSeek-V3~\cite{deepseek-isca} proved the great benefits from specialization focusing on a single LLM. However, as a software company they have to build upon existing hardware in the market.
The \emph{mortal computing} argument from Hinton~\cite{mortal} also partly parallels our vision on extremely specialized LPUs.

The first principle of DNN acceleration has been time-multiplexing hardware neurons and capturing data reuse, a \revadd{widely held} belief since DianNao in 2014~\cite{diannao}.
However, the data reuse chances are evaporating from modern LLM inference, which only has $\sim$1 operational intensity in the autoregressive decoding process.
Most researchers agree that external memory accesses are the key challenge, while having divergent \revadd{visions} of future.
Previous efforts focus on scheduling~\cite{yu2022orca, cai2025soma, mei2025helix, cheng2025concerto, sankaralingam2022mozart, stojkovic2025tapas}, quantization~\cite{darvish2023shared, liu2025vq, isscc2, gil2025avant, lascorz2024atalanta}, sparsity~\cite{gondimalla2023eureka, wang2024sofa, wang202228nm, zhao2024alisa}, speculative decoding~\cite{leviathan2023speculative}, and revisiting spiking neurons~\cite{isscc1},
but these techniques cannot fundamentally solve the memory access issue.
The Process-In-Memory (PIM) architecture has attracted considerable attention owing to its ability to reduce data movement and alleviate external memory accesses~\cite{asplos1,isscc3, heo2024neupims, gu2025pim, liu2025optipim}.
However, it cannot fundamentally eliminate the memory accesses required for weight loading, or suffers from high AD/DA overhead.

After the success of LLM, Language Processing Units (LPU) are emerging as the next important processor scheme.
Groq released the first commercialized LPU, rebranded from its Tensor Streaming Multiprocessor in 2022~\cite{groq-lpu2},
which attempts to address the issue by incorporating huge on-chip SRAM.
Etched Sohu~\cite{etched_sohu} is the most aggressively specialized LPU project that etched Transformer model architecture directly in the fabric.
However, these LPUs failed to etch weights, possibly due to the density and cost \revadd{limitations} explained in Section~\ref{sec::background}.
Proposed by this paper, we expect that Metal-Embedding will be the prerequisite technology for the emerging fully-specialized LPU products.

Metal-Embedding shares some underlying techniques with existing architectures:
Bespoke ML inference fabrics
Primitivization architecture proposed by Cambricon-C~\cite{cambricon-c},
\revtarget{cite2}\revadd{bespoke machine learning accelerators in printed and flexible electronics~\cite{mubarik2020printed,weller2021printed,ozer2023malodour,chakraborty2024hasics},}
bit-serial architecture explored by Stripes \cite{stripes},
and the prefabricated wafers once emerged as sea-of-gates architecture in 1990s~\cite{sea-of-gates}.

\section{Conclusions}
\label{sec::conclusion}

This paper introduces the Hardwired-Neurons Language Processing Unit (\uname{}), an extremely specialized processor that hardwires LLM weight parameters into its compute fabric. Enabled by a novel Metal-Embedding methodology that encodes weights in the 3D topology of metal wires, the design achieves a 15× density improvement and reduces photomask costs by 112×. The resulting system demonstrates unprecedented efficiency, delivering a 5,555× throughput gain and a 1,047× energy efficiency improvement over H100, establishing an economically viable, sustainable, ultra-high-performance cognitive substrate for general tasks.

%% file: tex/dataflow_appendix.tex
\section{Detailed Dataflow Description} \label{sec:appendix}

\subsection{Grouped-Query Attention}
\label{sec::gqa}
Figure~\ref{fig:dataflow}.(\uppercase\expandafter{\romannumeral 2}) illustrates the computation dataflow of the 64 query heads projection.
The token activation $X$ in the chip array has shape $(1, 2880)$. 
The activation $X$ in all chips is split into four equal slices of $(1, 720)$ and each chip takes one of \revadd{the} slices to calculate the partial sum of the query tensor.
Each chip contains a private, hard-wired slice of $W_q$ with shape $(720, 1024)$ inside the HN array. Therefore the product is generated locally and without any weight fetch. 
The four partial products are summed by a column-internal reduce operation, yielding \revadd{a} 16-head query vector $(1, 1024)$. The 16 query heads are then reshaped into $(2,8,64)$, \revadd{reflecting the} Grouped-Query Attention \revadd{structure where} every 8 query heads correspond to a single KV head.

\smallskip
Figure~\ref{fig:dataflow}.(\uppercase\expandafter{\romannumeral 3}) follows a similar spatial pattern for the key path. Each chip multiplies its $(1,720)$ input slice with its hard-wired $W_k$ slice $(720,128)$ and emits a partial key vector $(1,128)$. After the column-internal reduction and reshape, the new key head of $(2, 64)$ is \revadd{held} in chip\# (\# $=\ell\bmod 4$, where $\ell$ is the token’s position in the sequence). 
At this point the data layout is intentionally asymmetric: the query vector is fully replicated across columns, whereas each key vector is unique per chip in the \emph{sequence} dimension, a choice that minimizes the traffic of the \revadd{subsequent} dot-product. The data flow of $X \cdot W_{v}$ \revadd{mirrors} that of $W_k$.

\smallskip
Figure~\ref{fig:dataflow}.(\uppercase\expandafter{\romannumeral 4}) shows the computation of attention weight $Z$ in column-$i$ chips. Every chip already has the complete duplicate of the column-$i$ $Q$ heads $(2, 8, 64)$. The key tensor is partitioned horizontally, each chip keeps cached $K$ with shape $(2, 64,\ell/4)$ in its local KV-cache ($\ell$ is the current context length). The VEX unit multiplies $Q$ head $(2, 8, 64)$ with $K$ $(2, 64, \ell/4)$ and produces the local attention weight $Z$ $(2, 8, \ell/4)$. 
Because each chip sees only $\ell/4$ tokens, a column-wise all-reduce \revadd{needs to be performed}, after which every chip completes the normalization of its local fragment. 

\smallskip
Figure~\ref{fig:dataflow}.(\uppercase\expandafter{\romannumeral 5}) completes the attention score. $V$ is tiled exactly like $K$; each chip reads a $V$ slice $(2, \ell/4, 64)$ from its KV-cache. The VEX unit multiplies the $V$ $(2, 8, \ell/4)$ with the local attention weight $Z$ $(2, 8, \ell/4)$, and emits the partial-$O$ tensors $(2, 8, 64)$. 
A column-wise all-reduce needs to perform to add the four partial-$O$ tensors. After that, all chips in column-$i$ contain the 16 heads of the attention score $O$ with the shape $(2, 8, 64)$. Then, the matrix is flattened to the shape $(1, 1024)$ for the multiplication with $W_o$.

\smallskip
Figure~\ref{fig:dataflow}.(\uppercase\expandafter{\romannumeral 6}) depicts the output projection and first residual path. After the attention score computation, each of the four chips in \revadd{the} same column now holds the same, flattened attention score for 16 heads. The $W_o$ matrix is partitioned row-wise across the column group chips, with each column assigned a weight shape of $(1024, 2880)$ and each individual chip's HN array receiving a $(1024, 720)$. Each chip generates a partial-$O$ of shape $(1, 720)$. These partial-$O$s are combined via one row-wise all-reduce and one column-wise all-gather to yield the final $X_{o}$ of shape (1, 2880) in all chips.

\subsection{Feed-Forward Network with MoE}
\label{sec::ffn}

Figure~\ref{fig:dataflow}.(\uppercase\expandafter{\romannumeral 7}) shows the experts router stage. After the Group Query Attention computation, all chips in \uname{} contain the complete $(1, 2880)$ $X_o$ vector. These values are passed through RMSNorm before entering the routing layer.
As $W_{\text{rout}}$ only accounts for about $0.01\%$ of the total weights, we replicated a copy of the router weights on all 16 chips to avoid inter-chip data exchange.
After the computation of $X_{\text{norm}} \cdot W_{\text{rout}}$, each chip obtains complete $X_{\text{rout}}$. Next, top-k and softmax \revadd{operations} are performed.
The top-k result is used to generate a masked input tensor, $X$, with a shape of $(128,1,2880)$. In this tensor, the values for the top-k experts are set to those of $X_{\text{norm}}$, while all others are set to zero. Additionally, the top-k results are normalized with a softmax to obtain the expert weights.

\smallskip

Figure~\ref{fig:dataflow}.(\uppercase\expandafter{\romannumeral 8}) 
shows the up- and gate-projection stage.
Following the top-k masking, each chip processes its masked $X_{\text{mask}}$ tensor, which consists of 8 vectors, each of shape $(1,2880)$. Among the 128 $X_{\text{mask}}$ $(1, 2880)$ vectors in all chips, only $k$ are non-zero.
The weight of $W_{\text{up}}$ is sliced into 16 tiles, with the shape of $(8, 2880, 2880)$ for each chip.
Then, the local 8 vectors of $X_{\text{mask}}$ are multiplied with the sliced weight $W_{\text{up}}$ of eight experts. The 16 chips produce a total $X_{\text{up}}$ tensor of shape $(128,1,2880)$.
Because each chip stores the complete weight matrices for all experts, this step requires no inter-chip data communication.
The gate projection follows the same partitioning: $X_\text{{mask}}$ is multiplied by the corresponding $W_{\text{gate}}$ slice, yielding \revadd{an} $X_G$, $(128, 1, 2880)$ in total and $(8, 1, 2880)$ for each chip.
Applying the SwiGLU activation to $X_G$ and $X_\text{{up}}$, and taking the element-wise product, we get the output $X_t$ for the subsequent down-projection.

\smallskip
Figure~\ref{fig:dataflow}.\uppercase\expandafter{\romannumeral 9} shows the down-projection and second residual path. Still, $X_t$ in each chip with the shape of $(8,\,1,\,2880)$ multiplies its down weight slice $W_{\text{down}}$, $(8,\,2880,\,2880)$, to produce a partial $X_{\text{down}}$, $(8,\,1,\,2880)$. 
Next, the expert weights, which were obtained from the previous stage (\uppercase\expandafter{\romannumeral 7}), are multiplied with the corresponding expert outputs to get the weighted output for each expert.
Subsequently, \revadd{an} all-chip all-reduce operation is performed to sum the partial results. The shape of $X_{\text{down}}$ in all chips is from $(128,\,1,\,2880)$ to $(1,\,2880)$. The summed $X_{\text{down}}$ is then added to $X_o$ to yield the final layer output $Y\,(1,\,2880)$.

%% file: tex/cost_analysis.tex
\revtarget{appb}

{\section{Notes to Table~\ref{tab:tco_analysis_detailed}.}\label{sec:appendix2}
}
\revadd{
\noindent\textbf{\textsuperscript{1}Deployment scale.}
We define the "Low Volume" scenario as a single HNLPU system. The "High Volume" scenario targets OpenAI-scale throughput ($\sim$100\,M tokens/s~\cite{openai_devday, openai_nvidia_systems_partnership_2024}), corresponding to a 50-system HNLPU cluster. To ensure a fair TCO comparison, we normalize hardware counts based on equivalent inference throughput.
Under a high-concurrency workload (1K prefill/1K decode, concurrency 50), the average throughput per H100 GPU in a distributed setting is 1.08\,K tokens/s~\cite{clarifai2025b200h100}. Consequently, given the HNLPU's $\sim$2\,M tokens/s throughput under the same workload configuration, we equate one HNLPU system to approximately 2,000 H100 GPUs.
}

\revadd{
\noindent\textbf{\textsuperscript{2}Facility power modeling / PUE.}
Facility-level PUE is assumed to be 1.4, consistent with modern hyperscale AI datacenters~\cite{large-data-centers}.
}

\begin{table}[t]
  \centering
  \begin{threeparttable}
  \caption{\uname{} Cost Analysis.}
  \label{table:hn-cost-breakdown}
  \footnotesize
 {\begin{tabular*}{0.95\columnwidth}{@{\extracolsep{\fill}} l r } 
  \toprule
    - & Cost (\$) \\
  \midrule
    \textit{Recurring Cost (\$ / chip)} \\
    \quad Wafer            & 629 \\
    \quad Package \& Test   & 111 $\sim$ 185 \\
    \quad HBM        & 1,920 $\sim$ 3,840 \\
    \quad System Integration    & 1,900 $\sim$ 3,800 \\
    \midrule
    \textit{Non-recurring Cost (\$)} \\
    \quad Photomask    \\
    \quad\quad Homogeneous Mask & 13.85\,M $\sim$ 27.69\,M \\
    \quad\quad Metal-Embedding Mask & 18.46\,M $\sim$ 36.92\,M \\
    \quad Design \& Development     \\
    \quad\quad Architecture & 1.87\,M $\sim$ 3.74\,M \\
    \quad\quad Verification & 9.97\,M $\sim$ 19.93\,M \\
    \quad\quad Physical & 4.80\,M $\sim$ 14.41\,M  \\
    \quad\quad IP & 10.23\,M $\sim$ 20.46\,M \\
    \midrule
    \textit{Total Cost Scenarios (\$)} \\
    \quad Initial Build (Full NRE + Recurring) \\
    \quad\quad 1-\uname{}   & 59.25\,M $\sim$ 123.3\,M  \\
    \quad\quad 50-\uname{}  & 62.83\,M $\sim$ 129.9\,M  \\
    \quad Re-spin (Metal-Embedding Mask + Recurring) \\
    \quad\quad 1-\uname{}  & 18.53\,M $\sim$ 37.06\,M  \\
    \quad\quad 50-\uname{}  & 22.11\,M $\sim$ 43.68\,M  \\
  \bottomrule
\end{tabular*}}
\end{threeparttable}
\end{table}

\revadd{
\noindent\textbf{\textsuperscript{3}Node price of H100 and \uname{}.}For H100 price, each NVIDIA HGX H100 platform (8 GPUs/node) costs about \$\,320,000, including server, intra-node networking and 3-year hardware warranty~\cite{thinkmate_hgx, routerswitch_supermicro}. 
For \uname{}, we break down the node cost in Table~\ref{tab:tco_analysis_detailed} into recurring cost and non-recurring engineering (NRE) cost in Table~\ref{table:hn-cost-breakdown}.
}

\revadd{
Regarding recurring cost, we first estimate the silicon cost. 
Assuming a cost of \$\,16{,}988 for a 300\,mm 5\,nm wafer~\cite{cset_5nm_wafer_17000,Khan2020AIChips}, Murphy's model ($D_0=0.11$\,def/cm$^2$) predicts a 43\% yield ($\sim$27 of 62 dies), resulting in \(\approx \$629\) per good die.
Second, packaging and testing are estimated at \$3{,}000--\$5{,}000 per wafer to account for 2.5D integration complexity~\cite{imaps_2p5d_3d_cost}, resulting in an amortized cost of \$111--\$185 per chip.
Third, given an HBM cost of \$10--\$20/GB~\cite{eeworld2025hbm,depend_hbm_market_insight}, the 8-stack configuration (24\,GB per stack) amounts to \$1{,}920--\$3{,}840 per \uname{} module.
Finally, we include system integration costs, covering the chassis, motherboard, cooling, power, and CXL interconnects; 
these figures align with per-chip costs from established commercial platforms~\cite{semianalysis_memory_is_biggest_loser}.
}

\revadd{
Regarding one-time costs, we distinguish between photomask NRE and design \& development expenses.
First, we model the photomask NRE using a normalized cost model based on lithography complexity for the 5\,nm technology node. To account for the disparity in manufacturing costs, we assign a cost weighting factor of $6\times$ to EUV reticles relative to standard 193i DUV reticles~\cite{ebeam_economics_mask}. Given a typical 5\,nm layer stack comprising 12 EUV and 58 DUV layers~\cite{semitwiki_tsmc_n5,anysilicon_economics_asic,VASHISHTHA2022105481,Jones2020LithoVision}, the total mask set value corresponds to $58 + (12 \times 6) = 130$ normalized DUV units. 
In \uname{} architecture, the metal-embedding configuration requires 10 DUV reticles (VIA7, M8 Mandrel, M8 Cut, VIA8, M9 Mandrel, M9 Cut, VIA9, M10, VIA10, M11); consequently, this variable portion accounts for $7.7\%$ ($10/130$) of the full mask set, while the remaining $92.3\%$ represents the homogeneous mask cost shared across all variants. Anchoring the absolute 5\,nm mask set cost to a range of \$\,15\,M (\emph{optimistic}) to \$\,30\,M (\emph{pessimistic})~\cite{Graening_30M,dylan,EUVLitho2017MaskCost}, we derive a shared homogeneous mask cost of \$\,13.85--\$\,27.69\,M. The metal-embedding cost is estimated at \$\,1.15--\$\,2.31\,M per variant, amounting to \$\,18.46--\$\,36.92\,M in total for 16 chips.
Second, for \uname{} design and development costs, we derive our estimates from internal engineering data and design experience.
}

\revadd{
\noindent\textbf{\textsuperscript{4}Data center infrastructure.}
We consider two primary capital expenditures: inter-node networking and facility construction. 
For the H100 cluster baseline, we assume a standard three-tier non-blocking Fat-Tree topology.
In terms of hardware composition, the network fabric is built using NVIDIA ConnectX-7~\cite{connectx7_price} network interface cards (NICs) and Quantum-2 (QM9700) InfiniBand switches~\cite{fs_switch}, interconnected with corresponding optical transceivers~\cite{fs_optics}.
Accounting for the NICs, switches, and cabling costs, the estimated network equipment capital expenditure is approximately \$45\,K per node.
The facility construction cost is modeled as \$12\,M per MW of critical IT load~\cite{cushman_datacenter_cost}.
For \uname{}, we scale the networking cost based on the number of chips, while the construction cost is scaled based on total power consumption.
}

\revadd{
\noindent\textbf{\textsuperscript{5}Re-spin cost.}
For H100, changing the model does not require a re-spin; therefore, this cost is zero.
For HNLPU, the re-spin cost comprises the Non-Recurring cost for the metal-embedding mask, plus the recurring costs for fabricating, packaging, and testing, as listed in Table~\ref{table:hn-cost-breakdown}.
}

\revadd{
\noindent\textbf{\textsuperscript{6}Electricity.}
The electricity cost is calculated based on the industrial electricity price of \$0.095/kWh, representative of major U.S. data center hubs~\cite{eia_electricity}.
}

\revadd{
\noindent\textbf{\textsuperscript{7}Maintenance \& Support.}
For H100 clusters, Maintenance \& Support includes software licensing and hardware maintenance.
Software license fees are calculated based on NVIDIA AI Enterprise pricing guidelines~\cite{nvidia_ai_enterprise}.
Hardware maintenance is conservatively estimated as 5\% of hardware CapEx per year~\cite{Rasmussen_APC_WP6,Koomey2007SimpleModel}. 
For \uname{}, we model this cost by provisioning spare nodes: one for the low-volume scenario and five for the high-volume scenario.
}

\revadd{
\noindent\textbf{\textsuperscript{8}CO$_2$ emission.}
Our carbon emission analysis incorporates both embodied emissions from hardware manufacturing and operational emissions from energy consumption. We estimate the manufacturing emission for a single H100 card or an HNLPU module at 124.9\,kgCO$_2$e~\cite{zhao20243d, li2025ecoserve}. The grid carbon intensity is assumed to be 0.38\,kgCO$_2$e per kWh~\cite{gupta2022act}.
}